\documentclass{article}

\usepackage{microtype}
\usepackage{graphicx}
\usepackage{subcaption}
\usepackage{booktabs} 

\usepackage{hyperref}

\usepackage[preprint]{icml2026}

\usepackage{amsmath}
\usepackage{amssymb}
\usepackage{mathtools}
\usepackage{amsthm}

\usepackage[capitalize,noabbrev]{cleveref}

\theoremstyle{plain}

\theoremstyle{definition}

\theoremstyle{remark}

\usepackage{multirow}
\usepackage{multicol}
\usepackage[table]{xcolor}
\usepackage{pifont}

\usepackage[textsize=tiny]{todonotes}

\icmltitlerunning{Submission and Formatting Instructions for ICML 2026}

\begin{document}

\twocolumn[
  \icmltitle{AffectGPT-R1: Leveraging Reinforcement Learning for Open-Vocabulary Multimodal Emotion Recognition}



  \icmlsetsymbol{equal}{*}

  \begin{icmlauthorlist}
    \icmlauthor{Zheng Lian}{1,2}
    \icmlauthor{Fan Zhang}{3}
    \icmlauthor{Yazhou Zhang}{4}
    \icmlauthor{Jianhua Tao}{5}
    \icmlauthor{Rui Liu}{6}
    \icmlauthor{Haoyu Chen}{7}
    \icmlauthor{Xiaobai Li}{8}
    \icmlauthor{Bin He}{1,2}
  \end{icmlauthorlist}

  \icmlaffiliation{1}{State Key Laboratory of Autonomous Intelligent Unmanned Systems, Tongji University}
  \icmlaffiliation{2}{Frontiers Science Center for Intelligent Autonomous Systems, Ministry of Education, Tongji University}
  \icmlaffiliation{3}{The Chinese University of Hong Kong}
  \icmlaffiliation{4}{Tianjing University}
  \icmlaffiliation{5}{Department of Automation, BNRist, Tsinghua University}
  \icmlaffiliation{6}{Inner Mongolia University}
  \icmlaffiliation{7}{CMVS, University of Oulu}
  \icmlaffiliation{8}{Zhejiang University}
  
  \icmlcorrespondingauthor{Zheng Lian}{zheng.lian.zeroqiaoba@gmail.com}
  \icmlcorrespondingauthor{Jianhua Tao}{jhtao@tsinghua.edu.cn}

  \icmlkeywords{Machine Learning, ICML}

  \vskip 0.3in
]



\printAffiliationsAndNotice{}  

\begin{abstract}
Open-Vocabulary Multimodal Emotion Recognition (OV-MER) aims to predict emotions without being constrained by label spaces, enabling fine-grained emotion understanding. Unlike traditional discriminative methods, OV-MER leverages generative models to capture the full spectrum of emotions and employs emotion wheels (EWs) for metric calculation. Previous approaches (e.g., AffectGPT) primarily rely on token-level loss during training. However, this objective is misaligned with the metrics used in OV-MER, while these metrics cannot be optimized via gradient backpropagation. To address this limitation, we propose \textbf{AffectGPT-R1}, a reinforcement learning framework that treats EW-based metrics as a reward function and applies policy optimization to maximize this reward. Additionally, we introduce an explicit reasoning process and examine its necessity in OV-MER. To further guide model behavior, we incorporate auxiliary rewards that regularize both emotion reasoning and emotion prediction. We also apply length penalties to mitigate reward hacking. Experimental results demonstrate that AffectGPT-R1 yields significant performance improvements on OV-MER. Moreover, our approach enhances generalized emotion understanding, achieving state-of-the-art results on MER-UniBench. Our code is provided in the supplementary material and will be released to facilitate future research.
\end{abstract}

\section{Introduction}
\label{sec:intro}

Emotion is highly related to human cognition, decision-making, and behavior, playing a vital role in our daily life \cite{picard2000affective,xie2024emovit}. Naturally, humans express emotions through multiple modalities, including (micro)-gestures, (micro)-expressions, audio tones, linguistic content, and other cues \cite{ben2021video,chen2023smg,el2011survey,lian2026merbench}. To better understand human emotions, it is essential to integrate these clues, enabling models to capture emotional nuances. This has led to the development of Multimodal Emotion Recognition (MER) \cite{lian2023mer,zhang2025moda}. Traditional MER approaches primarily rely on Ekman’s theory \cite{ekman1992argument,li2023decoupled}, which categorizes human emotions into six basic labels: \emph{anger}, \emph{disgust}, \emph{happiness}, \emph{sadness}, \emph{fear}, and \emph{surprise}. However, human emotions extend far beyond these basic categories \cite{plutchik1980general,demszky2020goemotions}, and such a restricted label space inevitably leads to imprecise emotion descriptions, thereby impairing the emotional intelligence of developed systems.

Recently, open-vocabulary MER (OV-MER) has emerged as a promising research direction, aiming to shift emotion recognition from constrained categories to the full spectrum of emotion, thereby enabling more nuanced emotion representations \cite{lian2025affectgpt,lian2025ov}. To support this paradigm shift, OV-MER introduces new solutions and metrics. For solutions, it transitions from discriminative to generative approaches, leveraging the extensive vocabulary of large language models (LLMs) to expand the recognition scope of emotion labels. For metrics, it employs emotion wheels (EWs) to capture semantic relationships between distinct emotions. Appendix \ref{appendix:ew_metric} provides a detailed illustration of these metrics. To better address this task, current approaches primarily use token-level loss to align predicted and true labels \cite{lian2025affectgpt,cheng2024emotion}. However, these methods suffer from a critical misalignment: token-level loss shows limited correlation with EW-based metrics. For instance, \emph{sadness} and \emph{madness} exhibit high token-level similarity but low semantic similarity in their underlying emotions. Furthermore, EW-based metrics cannot be directly optimized via gradient backpropagation, introducing significant challenges and limitations to existing solutions.

In this paper, we propose \textbf{AffectGPT-R1}, which leverages EW-based metrics as the reward function and employs reinforcement learning to maximize this reward, enabling the direct optimization of these metrics. Meanwhile, we introduce an additional reasoning process and investigate its necessity in addressing OV-MER. To refine model behaviors, we incorporate auxiliary reward functions to constrain the model’s reasoning and emotion predictions. Additionally, we observe the issue of reward hacking \cite{everitt2017reinforcement}, wherein the model tends to produce lengthy predictions containing redundant emotional expressions. To address this, we introduce length penalties into the reward function, thereby encouraging more concise outputs while still maximizing overall reward. Experimental results demonstrate that AffectGPT-R1 achieves significant improvements on OV-MER. Our framework also attains state-of-the-art performance on MER-UniBench \cite{lian2025affectgpt}, a unified benchmark for generalized emotion understanding. Appendix \ref{appendix:novelty} provides a more detailed discussion of our novelty. The main contributions of this paper are summarized below:
\begin{itemize}

    \item \textbf{(Framework)} We introduce AffectGPT-R1, a pioneering work that applies reinforcement learning to OV-MER, shedding light on the role of reinforcement learning and reasoning mechanisms in this domain.

    \item \textbf{(Reward)} We systematically investigate various reward functions and conduct an in-depth analysis of their effectiveness. To mitigate reward hacking, we introduce length penalties, ensuring the model maximizes rewards while using the fewest emotional words.
    
    \item \textbf{(Performance)} We achieve state-of-the-art performance in open-vocabulary emotion detection and generalized emotion understanding, highlighting the potential of reinforcement learning in affective computing.

\end{itemize}

\section{Related Works}
\label{sec:formatting}

\subsection{Emotion Representation}
Emotion representation defines the recognition scope of developed models and serves as the foundation of MER \cite{gunes2011emotion}. There are two primary representation methods: dimensional emotion theory \cite{schlosberg1954three} and basic emotion theory \cite{ekman1992argument}.

\paragraph{Dimensional Emotion Theory.} 
This theory uses continuous dimensions to capture the core attributes underlying emotions. For example, Russell \cite{russell1980circumplex} proposed a two-dimensional model, characterizing emotions based on \emph{valence} (the pleasantness of an emotion) and \emph{arousal} (the activation level of an emotion). In addition, some researchers have introduced a third dimension, \emph{dominance}, to represent the degree of perceived control over emotions \cite{russell1977evidence}. However, these dimensions are abstract and less intuitive for the general public, thus posing challenges in practical applications.

\paragraph{Basic Emotion Theory.} 
This theory typically relies on six or eight basic labels to describe emotional states. For example, Ekman \cite{ekman1992argument} argued that human emotions can be categorized into six basic labels, whereas Plutchik \cite{plutchik1980general} identified eight primary emotions that form the basis for other emotional experiences. Nevertheless, human emotions are rich and diverse, extending far beyond these basic labels. Mapping such nuanced emotional expressions into a restricted set of basic labels inevitably leads to the loss of subtle emotional distinctions.

\paragraph{Open-vocabulary Emotion.}
To address these limitations, researchers propose an open-vocabulary representation manner \cite{lian2025ov}, which removes restrictions on the label space and enables models to predict any emotion. Unlike dimensional emotion theory, it focuses on more intuitive emotion words; Unlike basic emotion theory, it avoids restrictive label spaces, enabling more nuanced emotion representation. Therefore, the open-vocabulary paradigm opens up new opportunities for affective computing. However, this field is still in its early stages, and effective solutions remain unclear. This paper presents a pioneering effort that leverages reinforcement learning to tackle this task.

\begin{figure*}[t]
	\centering
	\includegraphics[width=\linewidth]{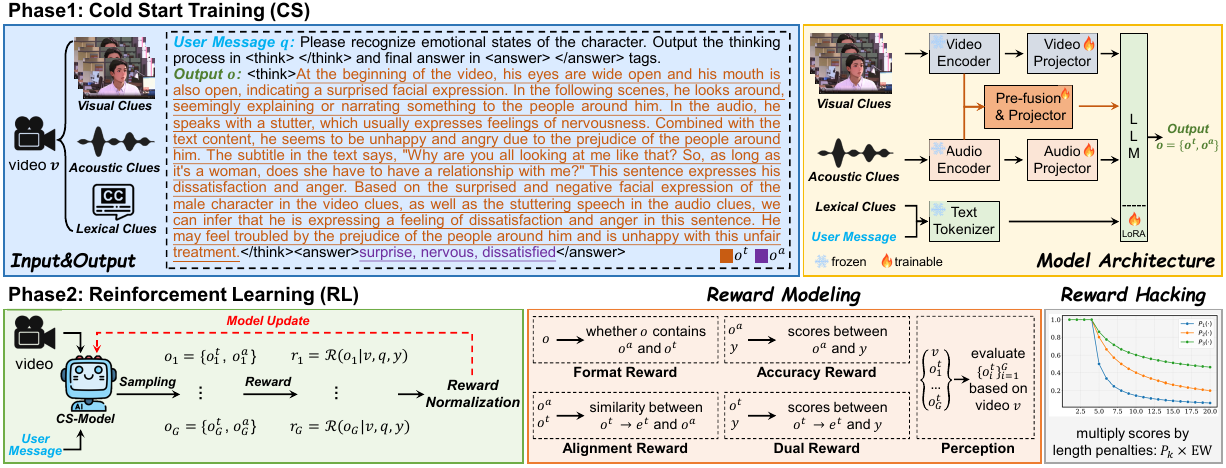}
	\caption{\textbf{Overall pipeline of AffectGPT-R1.} Our training pipeline consists of two phases: cold start training and reinforcement learning. In the first phase, we adopt the framework of AffectGPT but replace the generated content, including both thinking and answers. In the second phase, we propose five rewards and introduce length penalties to mitigate reward hacking.}
	\label{fig:pipeline}
\end{figure*}

\subsection{Reinforcement Learning}
Reinforcement learning has advanced the upper-bound performance of LLMs, demonstrating impressive results in tasks such as mathematical verification and code execution \cite{guo2025deepseek, shao2024deepseekmath, cobbe2021training}. This method primarily consists of two key components: policy optimization and reward modeling.

\paragraph{Policy Optimization.}
During training, we aim to maximize the reward function through policy updates. Proximal Policy Optimization (PPO) \cite{schulman2017proximal} is a widely used algorithm that leverages a value model for advantage estimation and incorporates token-wise KL penalties to prevent excessive optimization. Direct Preference Optimization (DPO) \cite{rafailov2023direct} focuses on directly optimizing policies to align with human preferences, eliminating the need for an explicit reward model and instead relying on an implicit reward model. This approach reduces computational costs and engineering complexity. Group Relative Policy Optimization (GRPO) \cite{shao2024deepseekmath} is a more efficient alternative to PPO, removing the value model and instead using a group-relative baseline computed from different outputs for the same input to guide optimization. Beyond these algorithms, some efforts have been made to enhance training efficiency and stability \cite{hu2025reinforce++, yu2025dapo}. This paper does not focus on designing new policy optimization algorithms, but rather on crafting optimal rewards for emotion recognition.

\paragraph{Reward Modeling.}  
Reward systems can be broadly categorized into rule-based and model-based rewards. The former is suitable for tasks with verifiable outcomes. For example, pass or fail unit tests in code generation \cite{le2022coderl}, output language requirements in question-answering tasks \cite{guo2025deepseek}, and IoU between predicted bounding boxes and ground-truth annotations in object detection \cite{shen2025vlm}. The latter is mainly used for tasks where rewards are difficult to compute, such as safety, helpfulness, and harmlessness \cite{zhang2024safetybench}. In prior work, how to design reward functions tailored for OV-MER remains unexplored. To this end, we introduce multiple rewards to supervise both thinking and answers, deriving insights into the effective rewards for this task.

\section{AffectGPT-R1}
\label{sec:affectgpt_r1}
Our training pipeline consists of two phases: cold start training and reinforcement learning. In the first phase, we utilize a \emph{large-scale, coarse-grained} dataset to establish initial capabilities in emotion understanding and format alignment. In the second phase, we use a \emph{high-quality, fine-grained} dataset for reward calculation and policy updates. Figure \ref{fig:pipeline} illustrates the overall pipeline of AffectGPT-R1.

\subsection{Notation Definition}
This section introduces some necessary notations. Specifically, we use $v$ to represent the video, which contains multimodal clues (such as visual, acoustic, and lexical information), and $q$ to denote the user message. The output $o$ consists of two components: the thinking $o^t$ and the answer $o^a$. Specifically, $o^t$ and $o^a$ correspond to the content enclosed within the $<$think$>$ and $<$/think$>$ tags and the $<$answer$>$ and $<$/answer$>$ tags, respectively. We denote the ground-truth open-vocabulary labels as $y$. Figure \ref{fig:pipeline} clarifies the meanings of these notations. Our main goal is to predict the emotion $y$ based on the video content $v$. We would like to emphasize that this paper focuses on utterance-level emotion recognition, ensuring fair comparison with prior work \cite{lian2025affectgpt,lian2025ov}. Appendix~\ref{appendix:task} provides additional clarification regarding our task formulation.

\subsection{Cold Start Training}
Human emotions are highly dependent on subtle cues \cite{li2022cas,li2022deep}, posing significant challenges for current models to capture all these signals and accurately interpret emotional states \cite{lian2024gpt,lu2024gpt}. To enhance emotion understanding, we employ \emph{large-scale coarse-grained} descriptive emotion datasets for cold-start training. In this phase, we adopt the same model architecture as AffectGPT \cite{lian2025affectgpt} (see Figure \ref{fig:pipeline}). This framework leverages a pre-fusion layer to effectively integrate multimodal cues, emphasizing the inherently multimodal nature of human emotions. Unlike the original AffectGPT, which only outputs emotion words, we generate both thinking (enclosed within $<$think$>$ and $<$/think$>$ tags) and answers (enclosed within $<$answer$>$ and $<$/answer$>$ tags). Since descriptive emotion datasets already follow a thinking-like format \cite{lian2023explainable}, we treat these descriptions as the thinking content. Given our primary focus on OV-MER, we use open-vocabulary emotions as the answer content. Figure \ref{fig:pipeline} illustrates an example of the input and output for cold-start training. This output format unifies descriptive and open-vocabulary emotions into a single task, enabling a single model to leverage their correlation and handle both tasks simultaneously. As demonstrated in Section \ref{sec:experiment_cold_start_data}, the quantity of cold-start data significantly affects both cold-start training performance and subsequent reinforcement learning. This further emphasizes our rationale for using \emph{large-scale} datasets in this phase.

\subsection{Reinforcement Learning}
\label{sec:reinforcement_learning}
Reinforcement learning relies on policy optimization and reward design. Rather than focusing on improving optimization strategies, we investigate how to design reward models that can better address OV-MER. In this section, we introduce various rewards to guide thinking and answers.

\subsubsection{Policy Optimization}
\label{sec:reinforcement_learning_policy}
We employ GRPO \cite{shao2024deepseekmath} for policy updates. Unlike PPO \cite{schulman2017proximal}, which relies on a critic model for advantage estimation, GRPO eliminates this component and instead uses group scores, significantly reducing memory overhead while improving training efficiency. Specifically, for a given input pair $(v,q)$, the policy model $\pi_{\theta_{\text{old}}}$ samples a group of responses $\mathbf{O}=\{o_i\}_{i=1}^{G}$. We then compute the reward for each response as $r_i = \mathcal{R}(o_i|v, q,y)$, where $\mathcal{R}(\cdot)$ is the reward function defined in Section \ref{sec:affectgpt_r1_reward}. To evaluate their relative quality, we normalize the rewards by computing their mean and standard deviation:
\begin{equation}
    \hat{A}_{i} = \frac{r_i - \text{mean}(\{r_i\}_{i=1}^G)}{\text{std}(\{r_i\}_{i=1}^G)}.
\end{equation}

During training, GRPO encourages the policy model $\pi_\theta$ to generate high-reward responses by minimizing the following objective function:
\begin{equation}
    \begin{aligned}
        \label{eq:grpo}
        \mathcal{J}_{\text{GRPO}}(\theta) =& \mathbb{E}_{(v,q) \sim \mathcal{D},\{o_i\}_{i=1}^G \sim \pi_{\theta_{\text{old}}}(\cdot|v,q)} \\
        &\left[\frac{1}{G}\sum_{i=1}^{G} \left( S_i
              - \beta {D}_\text{KL}\left(\pi_\theta||\pi_{\text{ref}}\right) \right)\right],
    \end{aligned}
\end{equation}
\begin{equation}
    S_i=\min \left(s_i\hat{A}_i, \text{clip}\left(s_i, 1-\epsilon, 1+\epsilon\right)\hat{A}_i\right),
\end{equation}
\begin{equation}
    s_i=\frac{\pi_{\theta}(o_i|v,q)}{\pi_{\theta_{\text{old}}}(o_i|v,q)},
\end{equation}
where $\pi_\theta$ represents the online policy model and $\pi_{\text{ref}}$ serves as the offline reference model. Similar to PPO, GRPO employs $\epsilon$ as the clipping threshold and incorporates a KL divergence penalty term to regularize the divergence between $\pi_\theta$ and $\pi_{\text{ref}}$. The combination of clipping and KL regularization effectively prevents extreme policy updates while maintaining training stability.

During experiments, we observed a divergence between the learned policy distribution and the initial model distribution, and the restrictive nature of the KL penalty term constrained overall performance. This trade-off between policy exploration and regularization has been proven in prior work \cite{yu2025dapo}. Following the prior work \cite{yu2025dapo}, we remove the KL regularization term and adopt the simplified objective function for policy update. Experimental results in Section \ref{sec:results_and_discussion} demonstrate the effectiveness of our method.
\begin{equation}
    \begin{aligned}
        \label{eq:grpo}
        \mathcal{J}_{\text{GRPO}}(\theta) = \mathbb{E}_{(v,q) \sim \mathcal{D},\{o_i\}_{i=1}^G \sim \pi_{\theta_{\text{old}}}(\cdot|v,q)} 
        \frac{1}{G}\sum_{i=1}^{G} S_i.
    \end{aligned}
\end{equation}

\subsubsection{Reward Modeling}
\label{sec:affectgpt_r1_reward}
Reward functions guide the update direction. This paper proposes five reward functions, all designed to steer thinking and answers. Section \ref{sec:experiment_reward} reveals the performance of the single reward and combinations of multiple rewards.

\paragraph{Format Reward.} 
To ensure structured outputs, we define a binary reward function that evaluates whether the model's response contains both thinking and answer components, returning 1 if compliant and 0 otherwise.
\begin{equation}
    \mathcal{R}_{\text{format}}(o|v, q, y) = \begin{cases}
        1, \mathrm{is\_required\_format}(o) \\
        0, \text{otherwise} \\
    \end{cases}
\end{equation}

\paragraph{Accuracy Reward.}
We employ the official metrics of OV-MER to compute accuracy rewards \cite{lian2025ov}. Unlike discriminative MER, which operates within a fixed label space, OV-MER imposes no constraints on the label space. Consequently, models are allowed to generate synonyms, i.e., emotions with identical meanings but different expressions, which poses challenges for evaluation. To mitigate these challenges, OV-MER introduces EW-based metrics, which leverage the structural information of emotion wheels for clustering. Following prior work \cite{lian2025ov}, we utilize five emotion wheels (see Appendix~\ref{appendix:ew}), and a detailed calculation procedure for EW-based metrics is presented in Appendix~\ref{appendix:ew_metric}. We denote this reward as:
\begin{equation}
    \mathcal{R}_{\text{accuracy}}(o|v, q, y) = \text{EW}\left(o^a, y\right).
\end{equation}

\paragraph{Alignment Reward.} \emph{Accuracy reward} ensures the correctness of $o^a$. However, without constraints on the reasoning process, the generated output may include irrelevant $o^a$ and $o^t$. To address this, we introduce \emph{alignment reward}, which ensures that the reasoning process is consistent with the final answer. Specifically, we extract emotion words $e^t$ from $o^t$ and compute their similarity with $o^a$. Appendix \ref{appendix:alignment} provides the detailed methods for emotion word extraction and similarity calculation. We denote this reward as:
\begin{equation}
    \mathcal{R}_{\text{alignment}}(o|v, q,y) = \mathrm{is\_similar}\left(o^t \rightarrow e^t, o^a\right).
\end{equation}

\paragraph{Dual Reward.} Similar to \emph{alignment reward}, this reward also imposes constraints on $o^t$. Specifically, we require the emotion words extracted from $o^t$ to match the true label $y$:
\begin{equation}
    \mathcal{R}_{\text{dual}}(o|v, q, y) = \text{EW}\left(o^t \rightarrow e^t, y\right).
\end{equation}

\paragraph{Perception Reward.}
Aforementioned rewards primarily evaluate text-level outputs $o$ but overlook whether the generated output aligns with the input video $v$. To assess perception correctness, inspired by EmoPrefer \cite{lian2026emoprefer}, we employ MLLMs to estimate pairwise preferences among outputs $(o_i, o_j)$ for a given video. Then, we apply the Bradley-Terry algorithm \cite{hunter2004mm} to derive the ranking from these preferences. Finally, we assign a reward of 1 to outputs ranked in the top 50\% and 0 to the rest. This reward is defined as follows:
\begin{equation}
    \mathcal{R}_{\text{perception}}(\{o_i\}_{i=1}^{G}|v,q,y) = \begin{cases}
        1, \mathrm{is\_top50}(o_i^t|v) \\
        0, \text{otherwise} \\
    \end{cases}
\end{equation}

\subsubsection{Reward Hacking}
\label{sec:method_reward_hacking}
During reinforcement learning, we observe that the model tends to generate lengthy outputs $o^a$ containing numerous emotional words. The primary reason for this is that, in the calculation of EW-based metrics, synonyms are automatically disregarded (see Appendix \ref{appendix:ew_metric}). Consequently, synonyms do not influence the metric scores, whereas predicting more words may instead increase the likelihood of matching the ground truth. This results in lengthy outputs, which in turn increases inference time and reduces the readability of $o^a$. This phenomenon, also referred to as reward hacking \cite{everitt2017reinforcement}, occurs when a model exploits loopholes or flaws in the reward function design to achieve high rewards while generating undesired behavior. To generate human-preferred outputs, we introduce length penalties based on the predicted label count, encouraging the model to maximize rewards with fewer prediction words. Specifically, we propose three types of penalties. Taking \emph{accuracy reward} as an example, these penalties are defined as follows:
\begin{align}
    P_{1}(o^a, y) &= \begin{cases}
        1, & |o^a| \leq |y| \\
        \frac{1}{|o^a| - |y| + 1}, & |o^a| > |y|
    \end{cases} \\
    P_2(o^a, y) &= \begin{cases}
        1, & |o^a| \leq |y| \\
        \frac{|y|}{|o^a|}, & |o^a| > |y|
    \end{cases} \\
    P_3(o^a, y) &= \begin{cases}
        1, & |o^a| \leq |y| \\
        \frac{\log(|y|)}{\log(|o^a|)}, & |o^a| > |y|
    \end{cases}
\end{align}
where $|o^a|$ and $|y|$ denote the number of emotion words in the predictions and ground truth, respectively. We then apply these length penalties to the accuracy reward, yielding the final accuracy score:
\begin{equation}
    \mathcal{R}_{\text{accuracy}}(o|v, q, y) = P_k(o^a, y) \times \text{EW}(o^a, y).
\end{equation}

Similar to the \emph{accuracy reward}, for the \emph{dual reward}, we also apply this penalty:
{\small
\begin{equation}
    \mathcal{R}_{\text{dual}}(o|v, q, y) = P_k(o^t \rightarrow e^t, y)\times\text{EW}(o^t \rightarrow e^t, y).
\end{equation}
}

Figure \ref{fig:pipeline} visualizes the values of these penalties (see the \emph{reward hacking} part). From $P_{1}(\cdot)$ to $P_{3}(\cdot)$, the penalty on prediction lengths becomes looser. In Section \ref{sec:experiment_penalty}, we conduct experiments to evaluate the effects of different penalty strategies on recognition performance and their efficacy in mitigating redundant predictions.

\section{Experimental Database and Setup}
\label{sec:experiment_setup}

\subsection{Corpus Description}
AffectGPT-R1 consists of two training phases: cold-start training and reinforcement learning. In the first phase, we utilize a \emph{large-scale} dataset to endow the model with emotion understanding and format alignment capabilities. For this purpose, we select MER-Caption+ \cite{lian2025affectgpt}, which contains 31K samples annotated with rich emotion descriptions and open-vocabulary labels. In the second phase, we require a \emph{high-quality} dataset specifically designed for OV-MER. Accordingly, we choose MER2025-OV \cite{lian2025mer}, which features human-annotated, high-quality open-vocabulary labels. For evaluation, we assess the model’s performance on OV-MERD+ \cite{lian2025affectgpt} and MER-UniBench \cite{lian2025affectgpt}, measuring its capabilities in open-vocabulary and generalized emotion understanding. We conduct a rigorous overlap check to ensure there are no overlapping samples between the training and testing data. More details are provided in Appendix~\ref{appendix:dataset}.

Table \ref{tab:dataset} presents statistics for these datasets. As shown, during cold-start training, we leverage both descriptive and open-vocabulary emotions provided by the dataset. However, in reinforcement learning, we exclusively use open-vocabulary emotions. This decision stems from the fact that the reward calculation process does not rely on human-annotated thinking results (see Section \ref{sec:affectgpt_r1_reward}). Therefore, our dataset selection is well-suited to the distinct requirements of each training stage.

\begin{table}[t]
	\centering
	\renewcommand\tabcolsep{3.2pt}
	\renewcommand\arraystretch{0.9}
	\caption{\textbf{Dataset statistics.} We summarize the datasets used for training and testing. In this table, ``Sen.'', ``Bas.'', ``OV'', and ``Des.'' are abbreviations for sentiment labels, basic emotions, open-vocabulary emotions, and descriptive emotions, respectively. We rigorously verify that there is no overlap between the training and testing data. More details can be found in Appendix~\ref{appendix:dataset}.}
    \label{tab:dataset}
    \scalebox{0.8}{
        \begin{tabular}{l|r|cccc}
            \toprule
            \multirow{2}{*}{\textbf{Dataset}} & \multirow{2}{*}{\textbf{\#Samples}} & \multicolumn{4}{c}{\textbf{Task}} \\
            & & Sen. & Bas. & OV & Des. \\
            \midrule
            \rowcolor{gray!20}
            \multicolumn{6}{c}{\emph{Training Phase 1: Cold-start training}} \\
            \midrule
            MER-Caption+ \cite{lian2025affectgpt} & 31,327 & $\times$ & $\times$ & $\surd$ & $\surd$ \\
            \midrule
            \rowcolor{gray!20}
            \multicolumn{6}{c}{\emph{Training Phase 2: Reinforcement learning}} \\
            \midrule
            MER2025-OV   \cite{lian2025mer}    & 1,000  & $\times$ & $\times$ & $\surd$ & $\times$ \\
            \midrule
            \rowcolor{gray!20}
            \multicolumn{6}{c}{\emph{Testing}} \\
            \midrule
            OV-MERD+     \cite{lian2025affectgpt}     & 532    & $\times$ & $\times$ & $\surd$ & $\times$ \\
            MER-UniBench \cite{lian2025affectgpt} & 12,799     & $\surd$ & $\surd$ & $\surd$ & $\times$ \\
            \bottomrule
        \end{tabular}
    }
\end{table}

\subsection{Implementation Details}
During training, we set the learning rate to 1e-5 and the maximum number of epochs to 60. The entire implementation is conducted using PyTorch, and the code is executed on an A100 GPU with 80 GB of memory. Due to memory constraints, we use a batch size of 3 for cold-start training and a batch size of 1 for reinforcement learning. For the model architecture, we adopt the default hyperparameters of AffectGPT \cite{lian2025affectgpt}. This setup ensures a fair comparison between AffectGPT and AffectGPT-R1, better highlighting the effectiveness of our proposed framework.

\section{Results and Discussion}
\label{sec:results_and_discussion}

\subsection{Reward Selection}
\label{sec:experiment_reward}
This section evaluates the effectiveness of different reward functions and identifies the optimal reward strategy for addressing OV-MER. For comparison, we report performance under three scenarios: (1) \emph{without reinforcement learning}, (2) \emph{with reinforcement learning using a single reward}, and (3) \emph{with reinforcement learning using multiple rewards}. Experimental results are presented in Table \ref{tab:reward}.

\paragraph{Single-reward RL.}
In the single-reward scenarios, we observe that the \emph{accuracy reward} and \emph{dual reward} are the most effective, significantly improving performance compared to the model without reinforcement learning. In contrast, other rewards lead to performance degradation, indicating potential reward hacking. The superior performance of the \emph{accuracy reward} and \emph{dual reward} stems from their direct optimization of EW-based metrics: one on $o^a$ (answer) and one on $o^t$ (thinking). This highlights the importance of directly aligning reinforcement learning with evaluation metrics in OV-MER. Meanwhile, the \emph{format reward} and \emph{alignment reward} show limited impact, because these aspects are already well-learned during the cold-start phase. The weak performance of the \emph{perception reward} may be due to current MLLMs' inability to accurately decode human preferences, which could introduce reward noise \cite{lian2026emoprefer}. Interestingly, the \emph{accuracy reward} and \emph{dual reward} yield similar performance, suggesting that while EW-based metrics are crucial, the choice of optimizing on thinking $o^t$ or answer $o^a$ has minimal impact. Therefore, in the next experiment, we use the \emph{accuracy reward} by default.

\paragraph{Multiple-reward RL.}
In the multiple-reward scenarios, we observe that combining the \emph{accuracy reward} and \emph{format reward} leads to further performance improvements compared to single-reward setups. However, using more rewards, such as all five simultaneously, results in performance degradation.
The primary issue arises from interference between rewards: combining multiple objectives can dilute their individual effectiveness, ultimately harming reinforcement learning outcomes. For instance, emphasizing one reward may inadvertently reduce the influence of others. These findings suggest that more rewards do not necessarily equate to better performance. Instead, selecting an appropriate combination is key. Based on our experimental results, the most effective reward setup is the combination of the \emph{accuracy reward} and \emph{format reward}.

\begin{table}[t]
	\centering
	\renewcommand\tabcolsep{6pt}
	\renewcommand\arraystretch{0.9}
	\caption{\textbf{Impact of reward function selection.} In this table, we report performance under three conditions: (1) \emph{without reinforcement learning}, (2) \emph{with reinforcement learning using a single reward}, and (3) \emph{with reinforcement learning using multiple rewards}.}
    \label{tab:reward}
    \scalebox{0.8}{
        \begin{tabular}{ccccc|c}
            \toprule
            \multicolumn{5}{c|}{\textbf{Reward Functions}} & \multirow{2}{*}{\textbf{OV-MERD+}} \\
            {accuracy} & {format} & {alignment} & {dual} & {perception} &  \\
            \midrule
            \rowcolor{gray!20}
            \multicolumn{6}{c}{\emph{Without reinforcement learning}} \\
            \midrule
            \ding{55} & \ding{55} & \ding{55} & \ding{55} & \ding{55} & 62.52 \\
            \midrule
            \rowcolor{gray!20}
            \multicolumn{6}{c}{\emph{With reinforcement learning (single reward)}} \\
            \midrule
            \ding{52} & \ding{55} & \ding{55} & \ding{55} & \ding{55} & 63.35\small\textcolor{teal}{${(+0.83)}$} \\
            \ding{55} & \ding{52} & \ding{55} & \ding{55} & \ding{55} & 59.22\small\textcolor{purple}{${(-3.30)}$} \\
            \ding{55} & \ding{55} & \ding{52} & \ding{55} & \ding{55} & 57.55\small\textcolor{purple}{${(-4.97)}$} \\
            \ding{55} & \ding{55} & \ding{55} & \ding{52} & \ding{55} & 64.98\small\textcolor{teal}{${(+2.46)}$} \\
            \ding{55} & \ding{55} & \ding{55} & \ding{55} & \ding{52} & 59.93\small\textcolor{purple}{${(-2.59)}$} \\
            \midrule
            \rowcolor{gray!20}
            \multicolumn{6}{c}{\emph{With reinforcement learning (multiple rewards)}} \\
            \midrule
            \ding{52} & \ding{52} & \ding{55} & \ding{55} & \ding{55} & \textbf{66.49}\small\textcolor{teal}{${(+3.97)}$} \\
            \ding{52} & \ding{52} & \ding{52} & \ding{55} & \ding{55} & 65.38\small\textcolor{teal}{${(+2.86)}$} \\
            \ding{52} & \ding{52} & \ding{52} & \ding{52} & \ding{55} & 64.66\small\textcolor{teal}{${(+2.14)}$} \\ 
            \ding{52} & \ding{52} & \ding{52} & \ding{52} & \ding{52} & 59.53\small\textcolor{purple}{${(-2.99)}$} \\
            \bottomrule
        \end{tabular}
    }
\end{table}

\subsection{Output Mode Analysis}
\label{sec:experiment_thinking}
In single-reward setups (see Table~\ref{tab:reward}), the most effective rewards are the \emph{accuracy reward} and \emph{dual reward}, both of which are directly optimized based on EW metrics. This observation motivates a heuristic hypothesis: \emph{Could we modify the output mode to generate and optimize solely on $o^a$, thereby encouraging the model to focus more on EW-based metrics?} Specifically, during cold-start training, we eliminate the thinking process and require the output to contain only $o^a$. During reinforcement learning, we optimize the \emph{accuracy reward} directly on $o^a$. We refer to the modified model as AffectGPT-R1 \emph{(w/o think)}, and the original model shown in Figure~\ref{fig:pipeline} as AffectGPT-R1 \emph{(w/ think)}. In Table~\ref{tab:think}, we observe that removing the thinking process yields improved performance. This improvement can be attributed to two factors. First, our thinking process is derived from coarse-grained datasets, where inherent noise may lead to flawed reasoning and consequently degrade performance. Second, incorporating additional thinking content distracts the model from solving OV-MER, thereby increasing training difficulty. Regardless of whether the thinking process proves useful, this work represents a pioneering work in applying reinforcement learning to OV-MER. Future work will involve designing high-quality cold-start data and conducting an in-depth analysis to determine which samples or emotional states require additional thinking processes.

\begin{table}[t]
	\centering
	\renewcommand\tabcolsep{10pt}
	\renewcommand\arraystretch{0.9}
	\caption{\textbf{Output mode impact}. This table reports the performance under different output types. We observe that a simpler model, one that only outputs answers and calculates rewards based solely on the answers, achieves better performance.}
    \label{tab:think}
    \scalebox{0.8}{
        \begin{tabular}{l|ccc}
            \toprule
            \textbf{Model} & \textbf{OV-MERD+} \\
            \midrule
            AffectGPT                 & 62.52 \\
            AffectGPT-R1 (w/ think)   & 66.49 \\ 
            AffectGPT-R1 (w/o think)  & \textbf{68.39} \\ 
            \bottomrule
        \end{tabular}
    }
\end{table}

\subsection{Impact of Different Penalties}
\label{sec:experiment_penalty}
During inference, the model tends to generate lengthy predictions containing redundant emotion words. This behavior stems from the suboptimal design of the reward function. To mitigate this issue, we introduce three types of penalties in Section \ref{sec:method_reward_hacking}, where the penalty strength increases from $P_3(\cdot)$ to $P_1(\cdot)$. In this section, we evaluate the effectiveness of these penalties across three dimensions: (1) \emph{their impact on performance}, (2) \emph{their ability to prevent overlong outputs}, and (3) \emph{whether they introduce reward degradation}. As shown in Table \ref{tab:penalty}, all three penalties successfully address overlong outputs without compromising performance. However, the strict penalty, $P_1(\cdot)$, leads to reward degradation. Therefore, $P_2(\cdot)$ and $P_3(\cdot)$ are better choices.

\begin{table}[h]
    \renewcommand\tabcolsep{10pt}
    \renewcommand\arraystretch{0.9}
	\centering
	\caption{\textbf{Impact of different penalties.} Our goal is to identify a penalty that maintains model performance, prevents overlong outputs, and avoids introducing reward degradation. Based on these criteria, $P_2(\cdot)$ and $P_3(\cdot)$ are more suitable choices.}
    \label{tab:penalty}
    \scalebox{0.8}{
        \begin{tabular}{c|c|c|c}
            \toprule
            \textbf{Penalty} & \textbf{Performance} & \textbf{Overlong} & \textbf{Degradation} \\
            \midrule
            $P_1(\cdot)$  & 67.44  & \ding{55}  & \ding{52} \\
            $P_2(\cdot)$  & 67.79  & \ding{55}  & \ding{55}  \\
            $P_3(\cdot)$  & 68.05  & \ding{55}  & \ding{55}  \\
            \midrule
            None   & 68.39  & \ding{52} & \ding{55}  \\
            \bottomrule
        \end{tabular}
    }
\end{table}

\begin{table*}[t]
    \centering
	\renewcommand\tabcolsep{3.96pt}
	\renewcommand\arraystretch{0.96}
	\caption{\textbf{Main results on MER-UniBench.} MER-UniBench is a unified benchmark comprising three key tasks: sentiment analysis, basic emotion recognition, and fine-grained emotion detection. We conduct a fair comparison among different baselines under identical input modalities. Overall, AffectGPT-R1 outperforms existing methods and achieves state-of-the-art performance in this benchmark.}
	\label{tab:main}
	\scalebox{0.8}{
		\begin{tabular}{l|cccc|cccc|c|c}
			\toprule
                &\multicolumn{4}{c|}{\textbf{Sentiment}} 
                &\multicolumn{4}{c|}{\textbf{Basic}} 
                &{\textbf{Fine-grained}} &\multirow{2}{*}{\textbf{Mean}} \\
                &MOSI &MOSEI &SIMS &SIMS v2 & MER2023 & MER2024 & MELD &IEMOCAP & OV-MERD+ & \\
			
            \midrule
        \rowcolor{gray!20}
        \multicolumn{11}{c}{\emph{Input Modality: Audio, Text}} \\
        \midrule

OneLLM \cite{han2024onellm} & 64.01 & 54.09 & 63.39 & 61.98 & 25.52 & 17.21 & 28.32 & 33.44 & 22.25 & 41.14 \\
SECap \cite{xu2024secap} & 55.76 & 54.18 & 59.51 & 57.41 & 40.95 & 52.46 & 25.56 & 36.92 & 36.97 & 46.64 \\
PandaGPT \cite{su2023pandagpt} & 66.06 & 61.33 & 62.93 & 58.88 & 33.57 & 39.04 & 31.91 & 36.55 & 31.33 & 46.84 \\
Qwen-Audio \cite{chu2023qwen} & 70.09 & 46.90 & 70.73 & 65.26 & 41.85 & 31.61 & 49.09 & 35.47 & 32.36 & 49.26 \\
SALMONN \cite{tang2023salmonn} & 81.00 & 67.03 & 68.69 & 65.93 & 55.53 & 45.38 & 45.62 & 46.84 & 45.00 & 57.89 \\
AffectGPT \cite{lian2025affectgpt} & \textbf{83.46} & \textbf{80.74} & 82.99 & \textbf{83.75} & 72.94 & 73.41 & \textbf{56.63} & 55.68 & 59.98 & 72.18 \\
\textbf{AffectGPT-R1 (Ours)} & 80.13 & 80.01 & \textbf{84.49} & 82.31 & \textbf{81.69} & \textbf{93.49} & \textbf{63.74} & \textbf{63.85} & \textbf{65.49} & \textbf{77.24} \\

        \midrule
        \rowcolor{gray!20}
        \multicolumn{11}{c}{\emph{Input Modality: Video, Text}} \\
        \midrule

Otter \cite{li2025otter} & 52.89 & 50.44 & 57.56 & 53.12 & 16.41 & 14.65 & 22.57 & 29.08 & 16.63 & 34.82 \\
Video-LLaVA \cite{lin2024video} & 56.37 & 61.64 & 53.28 & 57.45 & 36.93 & 30.25 & 30.73 & 38.95 & 34.00 & 44.40 \\
PandaGPT \cite{su2023pandagpt} & 58.50 & 64.25 & 62.07 & 65.25 & 39.13 & 47.16 & 38.33 & 47.21 & 35.07 & 50.77 \\
Video-ChatGPT \cite{maaz2024video} & 54.42 & 63.12 & 64.82 & 65.80 & 44.86 & 46.80 & 37.33 & 56.83 & 39.80 & 52.64 \\
VideoChat2 \cite{li2024mvbench} & 66.84 & 54.32 & 69.49 & 70.66 & 33.67 & 54.50 & 36.64 & 48.70 & 39.21 & 52.67 \\
LLaMA-VID \cite{li2024llama} & 61.78 & 63.89 & 69.35 & 67.48 & 50.72 & 57.60 & 42.75 & 46.02 & 45.01 & 56.07 \\
VideoChat \cite{li2025videochat} & 65.13 & 63.61 & 69.52 & 72.14 & 48.73 & 57.30 & 41.11 & 48.38 & 44.52 & 56.71 \\
Chat-UniVi \cite{jin2024chat} & 54.53 & 63.18 & 68.15 & 66.36 & 57.62 & 65.67 & 45.61 & 52.37 & 48.00 & 57.94 \\
mPLUG-Owl \cite{ye2023mplug} & 72.40 & 72.91 & 72.13 & 75.00 & 56.86 & 59.89 & 49.11 & 55.54 & 48.18 & 62.45 \\
AffectGPT \cite{lian2025affectgpt} & \textbf{82.39} & \textbf{81.57} & \textbf{87.20} & \textbf{86.29} & 74.58 & 75.29 & 57.63 & 62.19 & 61.65 & 74.31 \\
\textbf{AffectGPT-R1 (Ours)} & 78.78 & 79.07 & 85.91 & 85.85 & \textbf{77.72} & \textbf{85.29} & \textbf{61.09} & \textbf{67.42} & \textbf{62.42} & \textbf{75.95} \\

        \midrule
        \rowcolor{gray!20}
        \multicolumn{11}{c}{\emph{Input Modality: Audio, Video, Text}} \\
        \midrule

PandaGPT \cite{su2023pandagpt} & 61.92 & 67.61 & 68.38 & 67.23 & 40.21 & 51.89 & 37.88 & 44.04 & 37.12 & 52.92 \\
R1-Omni \cite{zhao2025r1} & 58.02& 56.48& 71.82& 68.58& 64.17& 67.43& 43.20& 51.58& 55.24& 59.61 \\
Emotion-LLaMA \cite{cheng2024emotion} & 66.13 & 67.66 & 78.32 & 77.23 & 59.38 & 73.62 & 46.76 & 55.47 & 52.97 & 64.17 \\
AffectGPT \cite{lian2025affectgpt} & \textbf{81.30} & \textbf{80.90} & \textbf{88.49} & \textbf{86.18} & 78.54 & 78.80 & 55.65 & 60.54 & 62.52 & 74.77 \\
\textbf{AffectGPT-R1 (Ours)} & 79.39 & 79.24 & 88.25 & 84.97 & \textbf{84.32} & \textbf{93.75} & \textbf{63.12} & \textbf{74.26} & \textbf{68.05} & \textbf{79.48} \\  

            \bottomrule
		\end{tabular}
	}
\end{table*}

\begin{figure*}[!t]
    \centering
    \includegraphics[width=\linewidth]{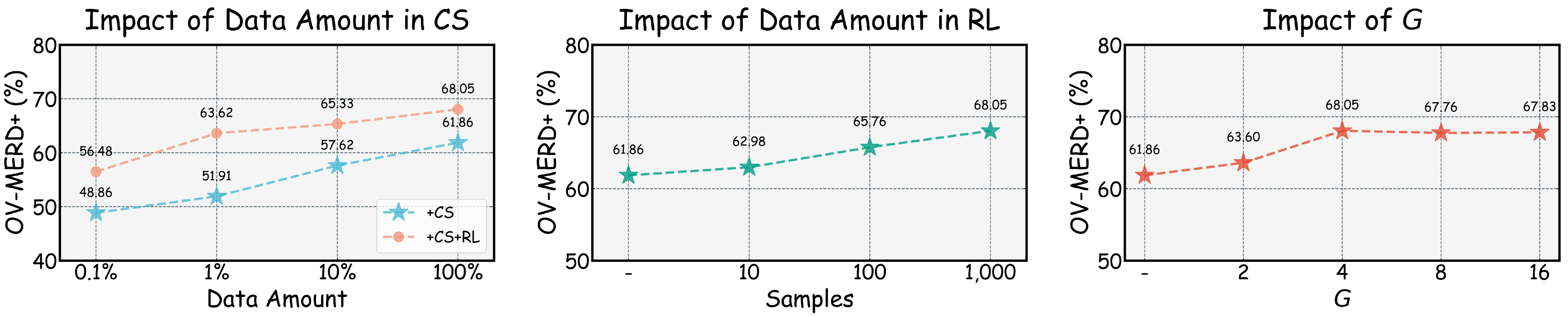}
    \caption{\textbf{Left: Impact of data amount in cold-start training.} This figure shows the performance of models trained with varying amounts of cold-start data and analyzes their influence on subsequent reinforcement learning. \textbf{Mid: Impact of data amount in reinforcement learning.} ``--'' denotes the baseline model without reinforcement learning. This figure shows that using a larger amount of data in reinforcement learning leads to better performance. \textbf{Right: Impact of $G$.} ``--'' denotes the baseline model without reinforcement learning.}
    \label{fig:ablation}
\end{figure*}

\subsection{Main Results on MER-UniBench}
\label{sec:experiment_main_results}
\paragraph{Global-level Analysis.}
The preceding sections focus on open-vocabulary emotion recognition. In this section, we extend our evaluation to MER-UniBench \cite{lian2025affectgpt}, a comprehensive benchmark that encompasses three tasks: sentiment analysis, basic emotion recognition, and fine-grained emotion detection. Appendix~\ref{appendix:dataset} provides additional details about this benchmark. Experimental results are summarized in Table \ref{tab:main}. To ensure a fair comparison, all methods are evaluated using the same input modality. Based on the average performance across three tasks, AffectGPT-R1 achieves state-of-the-art results on MER-UniBench, demonstrating the effectiveness of reinforcement learning in enhancing emotion understanding.

\paragraph{Task-level Analysis.}
Through further analysis of each task, AffectGPT-R1 demonstrates significant performance improvements in both basic and fine-grained emotion recognition, but shows a slight decline in sentiment analysis. This discrepancy stems from the fact that this work takes OV-MER as its primary focus and specifically tailors reward functions for this task. OV-MER aims to recognize discrete emotion words (e.g., \emph{happy} and \emph{surprise}), which closely align with the objectives of both basic and fine-grained emotion recognition. However, it differs somewhat from sentiment analysis, which targets emotional polarity (e.g., \emph{positive} and \emph{negative}). This distinction results in the reinforcement learning-optimized model, AffectGPT-R1, performing slightly worse on sentiment analysis compared to the non-optimized AffectGPT. Future work involves incorporating sentiment-related rewards into policy updates.

\subsection{Ablation Study}
\label{sec:ablation}



\paragraph{Data Amount in Cold-start Training.}
\label{sec:experiment_cold_start_data}

AffectGPT-R1 consists of two stages: cold-start training (denoted as $s_1$) and reinforcement learning (denoted as $s_2$). In this section, we examine the impact of the amount of cold-start data on both $s_1$ and $s_2$. Experimental results shown in Figure \ref{fig:ablation} (left) indicate that increasing the amount of cold-start data improves performance in $s_1$. Furthermore, the performance after reinforcement learning strongly correlates with the performance of the base model from $s_1$, and additional cold-start data also benefits $s_2$. This is primarily because reinforcement learning mainly reinforces correct responses rather than introducing new knowledge \cite{liu2025understanding}. Therefore, a high-quality base model is crucial for maximizing the effectiveness of reinforcement learning.


\paragraph{Data Amount in Reinforcement Learning.}
\label{sec:experiment_data_rl}
This section reveals the impact of data quantity on reinforcement learning. Specifically, starting from the same model obtained via cold-start training (i.e., the model trained on 100\% data, as shown in the left part of Figure \ref{fig:ablation}), we evaluate its performance using varying amounts of reinforcement learning data. Experimental results are shown in Figure \ref{fig:ablation} (mid). We observe that even using just 10 samples for reinforcement learning leads to performance improvements, confirming the effectiveness of reinforcement learning in addressing OV-MER. Moreover, increasing the amount of training data leads to greater performance gains. These phenomenon indicates that AffectGPT-R1 has not yet reached its upper-bound performance. In the future, we plan to collect more data to further improve the performance of AffectGPT-R1.



\paragraph{Impact of $G$.}
GRPO requires sampling $G$ outputs for group normalization. Figure \ref{fig:ablation} (right) reveals the impact of $G$. Due to GPU memory limitations, we restrict our comparisons to values of $G$ between 2 and 16 and do not evaluate larger values. Experimental results show that, regardless of the specific value of $G$, the model trained with reinforcement learning consistently outperforms the one without it, confirming the effectiveness of our approach. Meanwhile, the selection of $G$ also matters. A small $G$ does not fully exploit the benefits of GRPO, mainly because limited sampling reduces the chances of identifying optimal outcomes. In contrast, a larger $G$ (e.g., $G\geq4$) ensures the effectiveness of GRPO, but further increasing $G$ yields no significant improvement. Since a larger $G$ also increases computational cost, we set $G=4$ as the default configuration.

\section{Conclusion}
\label{sec:conclusion}
This paper proposes AffectGPT-R1, a pioneering work that applies reinforcement learning to OV-MER. We introduce various rewards and incorporate length penalties to mitigate reward hacking. Through extensive experiments, we reveal the impact of different rewards and identify the optimal reward design. AffectGPT-R1 shows strong effectiveness in OV-MER and also achieves state-of-the-art performance on MER-UniBench, a benchmark designed for generalized emotion understanding. Furthermore, we investigate the effects of various modules and hyperparameters, offering valuable insights for future research. By demonstrating the potential of reinforcement learning as a powerful paradigm for emotion understanding, our work will advance the development of this field toward a reinforcement learning era.




\section*{Impact Statement}
\paragraph{Social Impact.}
Emotion plays an important role in mental health, human-computer interaction, and embodied AI. This paper aims to advance the development of MER techniques, thereby promoting their applications.

\paragraph{Ethics Statement.}
We do not collect or annotate new data, but use existing datasets to verify the effectiveness of reinforcement learning in OV-MER. The use of all datasets has been authorized by their owners under corresponding usage requirements. Therefore, no ethical issues are raised.

\bibliography{mybib}

@article{shao2024deepseekmath,
  title={Deepseekmath: Pushing the limits of mathematical reasoning in open language models},
  author={Shao, Zhihong and Wang, Peiyi and Zhu, Qihao and Xu, Runxin and Song, Junxiao and Bi, Xiao and Zhang, Haowei and Zhang, Mingchuan and Li, YK and Wu, Yang and others},
  journal={arXiv preprint arXiv:2402.03300},
  year={2024}
}

@article{cobbe2021training,
  title={Training verifiers to solve math word problems},
  author={Cobbe, Karl and Kosaraju, Vineet and Bavarian, Mohammad and Chen, Mark and Jun, Heewoo and Kaiser, Lukasz and Plappert, Matthias and Tworek, Jerry and Hilton, Jacob and Nakano, Reiichiro and others},
  journal={arXiv preprint arXiv:2110.14168},
  year={2021}
}

@inproceedings{fang2025catch,
  title={Catch Your Emotion: Sharpening Emotion Perception in Multimodal Large Language Models},
  author={Fang, Yiyang and Liang, Jian and Huang, Wenke and Li, He and Su, Kehua and Ye, Mang},
  booktitle={Forty-second International Conference on Machine Learning},
  year={2025}
}

@article{zhao2025r1,
  title={R1-omni: Explainable omni-multimodal emotion recognition with reinforcement learning},
  author={Zhao, Jiaxing and Wei, Xihan and Bo, Liefeng},
  journal={arXiv preprint arXiv:2503.05379},
  year={2025}
}

@article{schulman2017proximal,
  title={Proximal policy optimization algorithms},
  author={Schulman, John and Wolski, Filip and Dhariwal, Prafulla and Radford, Alec and Klimov, Oleg},
  journal={arXiv preprint arXiv:1707.06347},
  year={2017}
}

@article{le2022coderl,
  title={Coderl: Mastering code generation through pretrained models and deep reinforcement learning},
  author={Le, Hung and Wang, Yue and Gotmare, Akhilesh Deepak and Savarese, Silvio and Hoi, Steven Chu Hong},
  journal={Advances in Neural Information Processing Systems},
  volume={35},
  pages={21314--21328},
  year={2022}
}

@article{rafailov2023direct,
  title={Direct preference optimization: Your language model is secretly a reward model},
  author={Rafailov, Rafael and Sharma, Archit and Mitchell, Eric and Manning, Christopher D and Ermon, Stefano and Finn, Chelsea},
  journal={Advances in Neural Information Processing Systems},
  volume={36},
  pages={53728--53741},
  year={2023}
}

@inproceedings{lian2025mer,
  title={Mer 2025: When affective computing meets large language models},
  author={Lian, Zheng and Liu, Rui and Xu, Kele and Liu, Bin and Liu, Xuefei and Zhang, Yazhou and Liu, Xin and Li, Yong and Cheng, Zebang and Zuo, Haolin and others},
  booktitle={Proceedings of the 33rd ACM International Conference on Multimedia},
  pages={13837--13842},
  year={2025}
}

@article{ekman1992argument,
  title={An argument for basic emotions},
  author={Ekman, Paul},
  journal={Cognition \& Emotion},
  volume={6},
  number={3-4},
  pages={169--200},
  year={1992},
  publisher={Taylor \& Francis}
}

@article{schlosberg1954three,
  title={Three dimensions of emotion.},
  author={Schlosberg, Harold},
  journal={Psychological Review},
  volume={61},
  number={2},
  pages={81},
  year={1954},
  publisher={American Psychological Association}
}

@incollection{plutchik1980general,
  title={A general psychoevolutionary theory of emotion},
  author={Plutchik, Robert},
  booktitle={Theories of Emotion},
  pages={3--33},
  year={1980},
  publisher={Elsevier}
}

@inproceedings{lu2024gpt,
  title={Gpt as psychologist? preliminary evaluations for gpt-4v on visual affective computing},
  author={Lu, Hao and Niu, Xuesong and Wang, Jiyao and Wang, Yin and Hu, Qingyong and Tang, Jiaqi and Zhang, Yuting and Yuan, Kaishen and Huang, Bin and Yu, Zitong and others},
  booktitle={Proceedings of the IEEE/CVF Conference on Computer Vision and Pattern Recognition},
  pages={322--331},
  year={2024}
}

@inproceedings{xu2024secap,
  title={Secap: Speech emotion captioning with large language model},
  author={Xu, Yaoxun and Chen, Hangting and Yu, Jianwei and Huang, Qiaochu and Wu, Zhiyong and Zhang, Shi-Xiong and Li, Guangzhi and Luo, Yi and Gu, Rongzhi},
  booktitle={Proceedings of the AAAI Conference on Artificial Intelligence},
  pages={19323--19331},
  year={2024}
}

@inproceedings{li2024llama,
  title={Llama-vid: An image is worth 2 tokens in large language models},
  author={Li, Yanwei and Wang, Chengyao and Jia, Jiaya},
  booktitle={European Conference on Computer Vision},
  pages={323--340},
  year={2024},
  organization={Springer}
}

@inproceedings{jin2024chat,
  title={Chat-univi: Unified visual representation empowers large language models with image and video understanding},
  author={Jin, Peng and Takanobu, Ryuichi and Zhang, Wancai and Cao, Xiaochun and Yuan, Li},
  booktitle={Proceedings of the IEEE/CVF Conference on Computer Vision and Pattern Recognition},
  pages={13700--13710},
  year={2024}
}

@inproceedings{han2024onellm,
  title={Onellm: One framework to align all modalities with language},
  author={Han, Jiaming and Gong, Kaixiong and Zhang, Yiyuan and Wang, Jiaqi and Zhang, Kaipeng and Lin, Dahua and Qiao, Yu and Gao, Peng and Yue, Xiangyu},
  booktitle={Proceedings of the IEEE/CVF Conference on Computer Vision and Pattern Recognition},
  pages={26584--26595},
  year={2024}
}

@inproceedings{lin2024video,
  title={Video-LLaVA: Learning United Visual Representation by Alignment Before Projection},
  author={Lin, Bin and Ye, Yang and Zhu, Bin and Cui, Jiaxi and Ning, Munan and Jin, Peng and Yuan, Li},
  booktitle={Proceedings of the 2024 Conference on Empirical Methods in Natural Language Processing},
  pages={5971--5984},
  year={2024}
}

@article{lian2024gpt,
  title={Gpt-4v with emotion: A zero-shot benchmark for generalized emotion recognition},
  author={Lian, Zheng and Sun, Licai and Sun, Haiyang and Chen, Kang and Wen, Zhuofan and Gu, Hao and Liu, Bin and Tao, Jianhua},
  journal={Information Fusion},
  volume={108},
  pages={102367},
  year={2024},
  publisher={Elsevier}
}

@article{el2011survey,
  title={Survey on speech emotion recognition: Features, classification schemes, and databases},
  author={El Ayadi, Moataz and Kamel, Mohamed S and Karray, Fakhri},
  journal={Pattern Recognition},
  volume={44},
  number={3},
  pages={572--587},
  year={2011},
  publisher={Elsevier}
}

@inproceedings{gunes2011emotion,
  title={Emotion representation, analysis and synthesis in continuous space: A survey},
  author={Gunes, Hatice and Schuller, Bj{\"o}rn and Pantic, Maja and Cowie, Roddy},
  booktitle={Proceedings of the International Conference on Automatic Face \& Gesture Recognition (FG)},
  pages={827--834},
  year={2011},
  organization={IEEE}
}

@inproceedings{demszky2020goemotions,
  title={GoEmotions: A Dataset of Fine-Grained Emotions},
  author={Demszky, Dorottya and Movshovitz-Attias, Dana and Ko, Jeongwoo and Cowen, Alan and Nemade, Gaurav and Ravi, Sujith},
  booktitle={Proceedings of the 58th Annual Meeting of the Association for Computational Linguistics},
  pages={4040--4054},
  year={2020}
}

@article{ben2021video,
  title={Video-based facial micro-expression analysis: A survey of datasets, features and algorithms},
  author={Ben, Xianye and Ren, Yi and Zhang, Junping and Wang, Su-Jing and Kpalma, Kidiyo and Meng, Weixiao and Liu, Yong-Jin},
  journal={IEEE Transactions on Pattern Analysis and Machine Intelligence},
  volume={44},
  number={9},
  pages={5826--5846},
  year={2021},
  publisher={IEEE}
}

@inproceedings{lian2024mer,
  title={Mer 2024: Semi-supervised learning, noise robustness, and open-vocabulary multimodal emotion recognition},
  author={Lian, Zheng and Sun, Haiyang and Sun, Licai and Wen, Zhuofan and Zhang, Siyuan and Chen, Shun and Gu, Hao and Zhao, Jinming and Ma, Ziyang and Chen, Xie and others},
  booktitle={Proceedings of the 2nd International Workshop on Multimodal and Responsible Affective Computing},
  pages={41--48},
  year={2024}
}

@article{chen2023smg,
  title={Smg: A micro-gesture dataset towards spontaneous body gestures for emotional stress state analysis},
  author={Chen, Haoyu and Shi, Henglin and Liu, Xin and Li, Xiaobai and Zhao, Guoying},
  journal={International Journal of Computer Vision},
  volume={131},
  number={6},
  pages={1346--1366},
  year={2023},
  publisher={Springer}
}

@inproceedings{zadeh2018multimodal,
  title={Multimodal language analysis in the wild: Cmu-mosei dataset and interpretable dynamic fusion graph},
  author={Zadeh, AmirAli Bagher and Liang, Paul Pu and Poria, Soujanya and Cambria, Erik and Morency, Louis-Philippe},
  booktitle={Proceedings of the 56th Annual Meeting of the Association for Computational Linguistics (Volume 1: Long Papers)},
  pages={2236--2246},
  year={2018}
}

@book{picard2000affective,
  title={Affective computing},
  author={Picard, Rosalind W},
  year={2000},
  publisher={MIT press}
}

@inproceedings{xie2024emovit,
  title={Emovit: Revolutionizing emotion insights with visual instruction tuning},
  author={Xie, Hongxia and Peng, Chu-Jun and Tseng, Yu-Wen and Chen, Hung-Jen and Hsu, Chan-Feng and Shuai, Hong-Han and Cheng, Wen-Huang},
  booktitle={Proceedings of the IEEE/CVF Conference on Computer Vision and Pattern Recognition},
  pages={26596--26605},
  year={2024}
}

@inproceedings{poria2019meld,
  title={Meld: A multimodal multi-party dataset for emotion recognition in conversations},
  author={Poria, Soujanya and Hazarika, Devamanyu and Majumder, Navonil and Naik, Gautam and Cambria, Erik and Mihalcea, Rada},
  booktitle={Proceedings of the 57th Conference of the Association for Computational Linguistics},
  pages={527--536},
  year={2019}
}

@inproceedings{zadeh2017tensor,
  title={Tensor fusion network for multimodal sentiment analysis},
  author={Zadeh, Amir and Chen, Minghai and Poria, Soujanya and Cambria, Erik and Morency, Louis-Philippe},
  booktitle={Proceedings of the Conference on Empirical Methods in Natural Language Processing},
  pages={1103--1114},
  year={2017}
}

@article{russell1980circumplex,
  title={A circumplex model of affect.},
  author={Russell, James A},
  journal={Journal of Personality and Social Psychology},
  volume={39},
  number={6},
  pages={1161},
  year={1980},
  publisher={American Psychological Association}
}

@article{russell1977evidence,
  title={Evidence for a three-factor theory of emotions},
  author={Russell, James A and Mehrabian, Albert},
  journal={Journal of Research in Personality},
  volume={11},
  number={3},
  pages={273--294},
  year={1977},
  publisher={Elsevier}
}

@article{guo2025deepseek,
  title={Deepseek-r1 incentivizes reasoning in llms through reinforcement learning},
  author={Guo, Daya and Yang, Dejian and Zhang, Haowei and Song, Junxiao and Wang, Peiyi and Zhu, Qihao and Xu, Runxin and Zhang, Ruoyu and Ma, Shirong and Bi, Xiao and others},
  journal={Nature},
  volume={645},
  number={8081},
  pages={633--638},
  year={2025},
  publisher={Nature Publishing Group UK London}
}

@article{liu2025understanding,
  title={Understanding r1-zero-like training: A critical perspective},
  author={Liu, Zichen and Chen, Changyu and Li, Wenjun and Qi, Penghui and Pang, Tianyu and Du, Chao and Lee, Wee Sun and Lin, Min},
  journal={arXiv preprint arXiv:2503.20783},
  year={2025}
}

@article{hu2025reinforce++,
  title={Reinforce++: A simple and efficient approach for aligning large language models},
  author={Hu, Jian},
  journal={arXiv preprint arXiv:2501.03262},
  year={2025}
}

@article{li2022deep,
  title={Deep learning for micro-expression recognition: A survey},
  author={Li, Yante and Wei, Jinsheng and Liu, Yang and Kauttonen, Janne and Zhao, Guoying},
  journal={IEEE Transactions on Affective Computing},
  volume={13},
  number={4},
  pages={2028--2046},
  year={2022},
  publisher={IEEE}
}

@article{li2022cas,
  title={CAS (ME) 3: A third generation facial spontaneous micro-expression database with depth information and high ecological validity},
  author={Li, Jingting and Dong, Zizhao and Lu, Shaoyuan and Wang, Su-Jing and Yan, Wen-Jing and Ma, Yinhuan and Liu, Ye and Huang, Changbing and Fu, Xiaolan},
  journal={IEEE Transactions on Pattern Analysis and Machine Intelligence},
  volume={45},
  number={3},
  pages={2782--2800},
  year={2022},
  publisher={IEEE}
}

@inproceedings{lian2026emoprefer,
  title={EmoPrefer: Can Large Language Models Understand Human Emotion Preferences?},
  author={Lian, Zheng and Sun, Licai and Chen, Lan and Chen, Haoyu and Cheng, Zebang and Zhang, Fan and Jia, Ziyu and Ma, Ziyang and Ma, Fei and Peng, Xiaojiang and others},
  booktitle={International Conference on Learning Representations},
  year={2026}
}

@inproceedings{yu2025dapo,
  title={Dapo: An open-source llm reinforcement learning system at scale},
  author={Yu, Qiying and Zhang, Zheng and Zhu, Ruofei and Yuan, Yufeng and Zuo, Xiaochen and Yue, Yu and Dai, Weinan and Fan, Tiantian and Liu, Gaohong and Liu, Lingjun and others},
  booktitle={Proceedings of the Advances in Neural Information Processing Systems},
  year={2025}
}

@inproceedings{zhang2024safetybench,
  title={SafetyBench: Evaluating the Safety of Large Language Models},
  author={Zhang, Zhexin and Lei, Leqi and Wu, Lindong and Sun, Rui and Huang, Yongkang and Long, Chong and Liu, Xiao and Lei, Xuanyu and Tang, Jie and Huang, Minlie},
  booktitle={Proceedings of the 62nd Annual Meeting of the Association for Computational Linguistics (Volume 1: Long Papers)},
  pages={15537--15553},
  year={2024}
}

@inproceedings{everitt2017reinforcement,
  title={Reinforcement learning with a corrupted reward channel},
  author={Everitt, Tom and Krakovna, Victoria and Orseau, Laurent and Legg, Shane},
  booktitle={Proceedings of the 26th International Joint Conference on Artificial Intelligence},
  pages={4705--4713},
  year={2017}
}

@article{shen2025vlm,
  title={Vlm-r1: A stable and generalizable r1-style large vision-language model},
  author={Shen, Haozhan and Liu, Peng and Li, Jingcheng and Fang, Chunxin and Ma, Yibo and Liao, Jiajia and Shen, Qiaoli and Zhang, Zilun and Zhao, Kangjia and Zhang, Qianqian and others},
  journal={arXiv preprint arXiv:2504.07615},
  year={2025}
}

@article{busso2008iemocap,
  title={IEMOCAP: Interactive emotional dyadic motion capture database},
  author={Busso, Carlos and Bulut, Murtaza and Lee, Chi-Chun and Kazemzadeh, Abe and Mower, Emily and Kim, Samuel and Chang, Jeannette N and Lee, Sungbok and Narayanan, Shrikanth S},
  journal={Language Resources and Evaluation},
  volume={42},
  pages={335--359},
  year={2008},
  publisher={Springer}
}

@inproceedings{liu2022make,
  title={Make acoustic and visual cues matter: Ch-sims v2. 0 dataset and av-mixup consistent module},
  author={Liu, Yihe and Yuan, Ziqi and Mao, Huisheng and Liang, Zhiyun and Yang, Wanqiuyue and Qiu, Yuanzhe and Cheng, Tie and Li, Xiaoteng and Xu, Hua and Gao, Kai},
  booktitle={Proceedings of the International Conference on Multimodal Interaction},
  pages={247--258},
  year={2022}
}

@inproceedings{lian2023mer,
  title={Mer 2023: Multi-label learning, modality robustness, and semi-supervised learning},
  author={Lian, Zheng and Sun, Haiyang and Sun, Licai and Chen, Kang and Xu, Mngyu and Wang, Kexin and Xu, Ke and He, Yu and Li, Ying and Zhao, Jinming and others},
  booktitle={Proceedings of the 31st ACM International Conference on Multimedia},
  pages={9610--9614},
  year={2023}
}

@article{li2025videochat,
  title={Videochat: Chat-centric video understanding},
  author={Li, KunChang and He, Yinan and Wang, Yi and Li, Yizhuo and Wang, Wenhai and Luo, Ping and Wang, Yali and Wang, Limin and Qiao, Yu},
  journal={Science China Information Sciences},
  volume={68},
  number={10},
  pages={200102},
  year={2025},
  publisher={Springer}
}

@inproceedings{su2023pandagpt,
  title={PandaGPT: One Model To Instruction-Follow Them All},
  author={Su, Yixuan and Lan, Tian and Li, Huayang and Xu, Jialu and Wang, Yan and Cai, Deng},
  booktitle={Proceedings of the 1st Workshop on Taming Large Language Models: Controllability in the era of Interactive Assistants},
  pages={11--23},
  year={2023}
}

@inproceedings{maaz2024video,
  title={Video-ChatGPT: Towards Detailed Video Understanding via Large Vision and Language Models},
  author={Maaz, Muhammad and Rasheed, Hanoona and Khan, Salman and Khan, Fahad},
  booktitle={Proceedings of the 62nd Annual Meeting of the Association for Computational Linguistics (Volume 1: Long Papers)},
  pages={12585--12602},
  year={2024}
}

@inproceedings{girdhar2023imagebind,
  title={Imagebind: One embedding space to bind them all},
  author={Girdhar, Rohit and El-Nouby, Alaaeldin and Liu, Zhuang and Singh, Mannat and Alwala, Kalyan Vasudev and Joulin, Armand and Misra, Ishan},
  booktitle={Proceedings of the IEEE/CVF Conference on Computer Vision and Pattern Recognition},
  pages={15180--15190},
  year={2023}
}

@inproceedings{li2024mvbench,
  title={MVBench: A Comprehensive Multi-modal Video Understanding Benchmark},
  author={Kunchang Li and Yali Wang and Yinan He and Yizhuo Li and Yi Wang and Yi Liu and Zun Wang and Jilan Xu and Guo Chen and Ping Luo and Limin Wang and Yu Qiao},
  booktitle={Proceedings of the IEEE/CVF Conference on Computer Vision and Pattern Recognition},
  year={2024}
}

@inproceedings{tang2023salmonn,
  title={SALMONN: Towards Generic Hearing Abilities for Large Language Models},
  author={Tang, Changli and Yu, Wenyi and Sun, Guangzhi and Chen, Xianzhao and Tan, Tian and Li, Wei and Lu, Lu and MA, Zejun and Zhang, Chao},
  booktitle={International Conference on Learning Representations},
  year={2023}
}

@article{ye2023mplug,
  title={mPLUG-Owl: Modularization Empowers Large Language Models with Multimodality},
  author={Ye, Qinghao and Xu, Haiyang and Xu, Guohai and Ye, Jiabo and Yan, Ming and Zhou, Yiyang and Wang, Junyang and Hu, Anwen and Shi, Pengcheng and Shi, Yaya and others},
  journal={arXiv preprint arXiv:2304.14178},
  year={2023}
}

@article{cheng2024emotion,
  title={Emotion-llama: Multimodal emotion recognition and reasoning with instruction tuning},
  author={Cheng, Zebang and Cheng, Zhi-Qi and He, Jun-Yan and Wang, Kai and Lin, Yuxiang and Lian, Zheng and Peng, Xiaojiang and Hauptmann, Alexander},
  journal={Advances in Neural Information Processing Systems},
  volume={37},
  pages={110805--110853},
  year={2024}
}

@inproceedings{yu2020ch,
  title={Ch-sims: A chinese multimodal sentiment analysis dataset with fine-grained annotation of modality},
  author={Yu, Wenmeng and Xu, Hua and Meng, Fanyang and Zhu, Yilin and Ma, Yixiao and Wu, Jiele and Zou, Jiyun and Yang, Kaicheng},
  booktitle={Proceedings of the 58th Annual Meeting of the Association for Computational Linguistics},
  pages={3718--3727},
  year={2020}
}

@article{awadalla2023openflamingo,
  title={Openflamingo: An open-source framework for training large autoregressive vision-language models},
  author={Awadalla, Anas and Gao, Irena and Gardner, Josh and Hessel, Jack and Hanafy, Yusuf and Zhu, Wanrong and Marathe, Kalyani and Bitton, Yonatan and Gadre, Samir and Sagawa, Shiori and others},
  journal={arXiv preprint arXiv:2308.01390},
  year={2023}
}

@inproceedings{chen2023beats,
  title={BEATs: audio pre-training with acoustic tokenizers},
  author={Chen, Sanyuan and Wu, Yu and Wang, Chengyi and Liu, Shujie and Tompkins, Daniel and Chen, Zhuo and Che, Wanxiang and Yu, Xiangzhan and Wei, Furu},
  booktitle={Proceedings of the 40th International Conference on Machine Learning},
  pages={5178--5193},
  year={2023}
}

@inproceedings{radford2023robust,
  title={Robust speech recognition via large-scale weak supervision},
  author={Radford, Alec and Kim, Jong Wook and Xu, Tao and Brockman, Greg and McLeavey, Christine and Sutskever, Ilya},
  booktitle={International Conference on Machine Learning},
  pages={28492--28518},
  year={2023},
  organization={PMLR}
}

@article{li2025otter,
  title={Otter: A multi-modal model with in-context instruction tuning},
  author={Li, Bo and Zhang, Yuanhan and Chen, Liangyu and Wang, Jinghao and Pu, Fanyi and Cahyono, Joshua Adrian and Yang, Jingkang and Li, Chunyuan and Liu, Ziwei},
  journal={IEEE Transactions on Pattern Analysis and Machine Intelligence},
  year={2025},
  publisher={IEEE}
}

@article{chu2023qwen,
  title={Qwen-audio: Advancing universal audio understanding via unified large-scale audio-language models},
  author={Chu, Yunfei and Xu, Jin and Zhou, Xiaohuan and Yang, Qian and Zhang, Shiliang and Yan, Zhijie and Zhou, Chang and Zhou, Jingren},
  journal={arXiv preprint arXiv:2311.07919},
  year={2023}
}

@article{lian2026merbench,
  title={Merbench: A unified evaluation benchmark for multimodal emotion recognition},
  author={Lian, Zheng and Sun, Licai and Ren, Yong and Gu, Hao and Sun, Haiyang and Chen, Lan and Liu, Bin and Tao, Jianhua},
  journal={IEEE Transactions on Pattern Analysis and Machine Intelligence},
  year={2026},
  publisher={IEEE}
}

@article{poria2019emotion,
  title={Emotion recognition in conversation: Research challenges, datasets, and recent advances},
  author={Poria, Soujanya and Majumder, Navonil and Mihalcea, Rada and Hovy, Eduard},
  journal={IEEE Access},
  volume={7},
  pages={100943--100953},
  year={2019},
  publisher={IEEE}
}

@article{sun2024svfap,
  title={Svfap: Self-supervised video facial affect perceiver},
  author={Sun, Licai and Lian, Zheng and Wang, Kexin and He, Yu and Xu, Mingyu and Sun, Haiyang and Liu, Bin and Tao, Jianhua},
  journal={IEEE Transactions on Affective Computing},
  year={2024},
  publisher={IEEE}
}

@article{lian2023explainable,
  title={Explainable multimodal emotion reasoning},
  author={Lian, Zheng and Sun, Licai and Xu, Mingyu and Sun, Haiyang and Xu, Ke and Wen, Zhuofan and Chen, Shun and Liu, Bin and Tao, Jianhua},
  journal={arXiv preprint arXiv:2306.15401},
  year={2023}
}

@inproceedings{lian2025ov,
  title={OV-MER: Towards Open-Vocabulary Multimodal Emotion Recognition},
  author={Lian, Zheng and Sun, Haiyang and Sun, Licai and Chen, Haoyu and Chen, Lan and Gu, Hao and Wen, Zhuofan and Chen, Shun and Siyuan, Zhang and Yao, Hailiang and others},
  booktitle={Forty-second International Conference on Machine Learning},
  year={2025}
}

@article{hunter2004mm,
  title={MM algorithms for generalized Bradley-Terry models},
  author={Hunter, David R},
  journal={Annals of Statistics},
  volume={32},
  number={1},
  pages={384--406},
  year={2004},
  publisher={Institute of Mathematical Statistics}
}

@inproceedings{lian2025affectgpt,
  title={AffectGPT: A New Dataset, Model, and Benchmark for Emotion Understanding with Multimodal Large Language Models},
  author={Lian, Zheng and Chen, Haoyu and Chen, Lan and Sun, Haiyang and Sun, Licai and Ren, Yong and Cheng, Zebang and Liu, Bin and Liu, Rui and Peng, Xiaojiang and others},
  booktitle={Forty-second International Conference on Machine Learning},
  year={2025}
}

@article{wang2023incomplete,
  title={Incomplete multimodality-diffused emotion recognition},
  author={Wang, Yuanzhi and Li, Yong and Cui, Zhen},
  journal={Advances in Neural Information Processing Systems},
  volume={36},
  pages={17117--17128},
  year={2023}
}

@inproceedings{zhang2025moda,
  title={Moda: Modular duplex attention for multimodal perception, cognition, and emotion understanding},
  author={Zhang, Zhicheng and Xia, Wuyou and Zhao, Chenxi and Yan, Zhou and Liu, Xiaoqiang and Zhu, Yongjie and Qin, Wenyu and Wan, Pengfei and Zhang, Di and Yang, Jufeng},
  booktitle={Forty-second International Conference on Machine Learning},
  year={2025}
}

@inproceedings{zhang2025videmo,
  title={VidEmo: Affective-Tree Reasoning for Emotion-Centric Video Foundation Models},
  author={Zhang, Zhicheng and Wang, Weicheng and Zhu, Yongjie and Qin, Wenyu and Wan, Pengfei and ZHANG, Di and Yang, Jufeng},
  booktitle={Thirty-ninth Annual Conference on Neural Information Processing Systems},
  year={2025}
}

@inproceedings{li2023decoupled,
  title={Decoupled multimodal distilling for emotion recognition},
  author={Li, Yong and Wang, Yuanzhi and Cui, Zhen},
  booktitle={Proceedings of the IEEE/CVF Conference on Computer Vision and Pattern Recognition},
  pages={6631--6640},
  year={2023}
}

@inproceedings{poria2017context,
  title={Context-dependent sentiment analysis in user-generated videos},
  author={Poria, Soujanya and Cambria, Erik and Hazarika, Devamanyu and Majumder, Navonil and Zadeh, Amir and Morency, Louis-Philippe},
  booktitle={Proceedings of the 55th annual meeting of the association for computational linguistics (volume 1: Long papers)},
  pages={873--883},
  year={2017}
}

@inproceedings{tsai2019multimodal,
  title={Multimodal Transformer for Unaligned Multimodal Language Sequences},
  author={Tsai, Yao-Hung Hubert and Bai, Shaojie and Liang, Paul Pu and Kolter, J Zico and Morency, Louis-Philippe and Salakhutdinov, Ruslan},
  booktitle={Proceedings of the 57th Annual Meeting of the Association for Computational Linguistics},
  pages={6558--6569},
  year={2019}
}
\bibliographystyle{icml2026}

\newpage
\appendix
\onecolumn

\section{Reproducibility Statement}
\label{appendix:reproducibility_statement}
In this paper, we have made every effort to ensure the reproducibility of our work. In the supplementary material, we provide our source code and command lines for cold-start training, reinforcement learning, and inference. Additionally, we include a detailed README file. To comply with ICML's anonymity requirements, we have removed all external links and owner names. After acceptance, we will release the full code on GitHub to facilitate further research.

\section{Novelty Statement}
\label{appendix:novelty}
In this paper, we present pioneering work that reveals the potential of reinforcement learning in OV-MER. There are currently two main types of emotion recognition tasks: \emph{discriminative emotions} and \emph{generative emotions}. The former primarily involves tasks with a constrained label space, such as basic or dimensional emotions \cite{wang2023incomplete,sun2024svfap,fang2025catch}. In contrast, the latter imposes no restrictions on the label space, giving rise to the emerging area of open-vocabulary emotions \cite{lian2025ov}. Recent research has shown a growing interest in shifting from \emph{discriminative emotions} to \emph{generative emotions}, driven by the need for more nuanced emotional representations \cite{lian2025affectgpt,cheng2024emotion,zhang2025videmo}. However, effectively addressing \emph{generative emotions} remains an open challenge. In this work, we focus on open-vocabulary emotions, a specific subtask within \emph{generative emotions}, and propose leveraging reinforcement learning as a solution approach. During our investigation, several practical issues emerged, including the design of reward functions and strategies to mitigate reward hacking. By tackling these challenges, our experimental results demonstrate the effectiveness of reinforcement learning in handling \emph{generative emotions}. Therefore, this paper should not be viewed merely as an application of reinforcement learning, but rather as a significant milestone in advancing \emph{generative emotions} research. This work will inspire more researchers to consider reinforcement learning as a foundational tool for addressing \emph{generative emotions}.

Meanwhile, we would like to emphasize that this work differs from R1-Omni \cite{zhao2025r1}, another RL-based framework in affective computing. Specifically, R1-Omni focuses on \emph{discriminative emotions}, whereas AffectGPT-R1 shifts the task to \emph{generative emotions}. Due to this task difference, we adopt distinct reward functions and face unique challenges, such as reward hacking, that are not present in R1-Omni. Meanwhile, it is important to note that R1-Omni is only available as a preprint on arXiv. Following the ICML policy, we are not required to compare with it. Nevertheless, to thoroughly validate the effectiveness of AffectGPT-R1, we include a comparison with R1-Omni, and the experimental results demonstrate our model’s advantages in open-vocabulary emotion detection and generalized emotion understanding.

\section{Emotion Wheel (EW)}
\label{appendix:ew}
\begin{figure}[h]
    \begin{center}
        \begin{subcaptionbox}{W1}[0.26\linewidth]
            {\includegraphics[width=\linewidth]{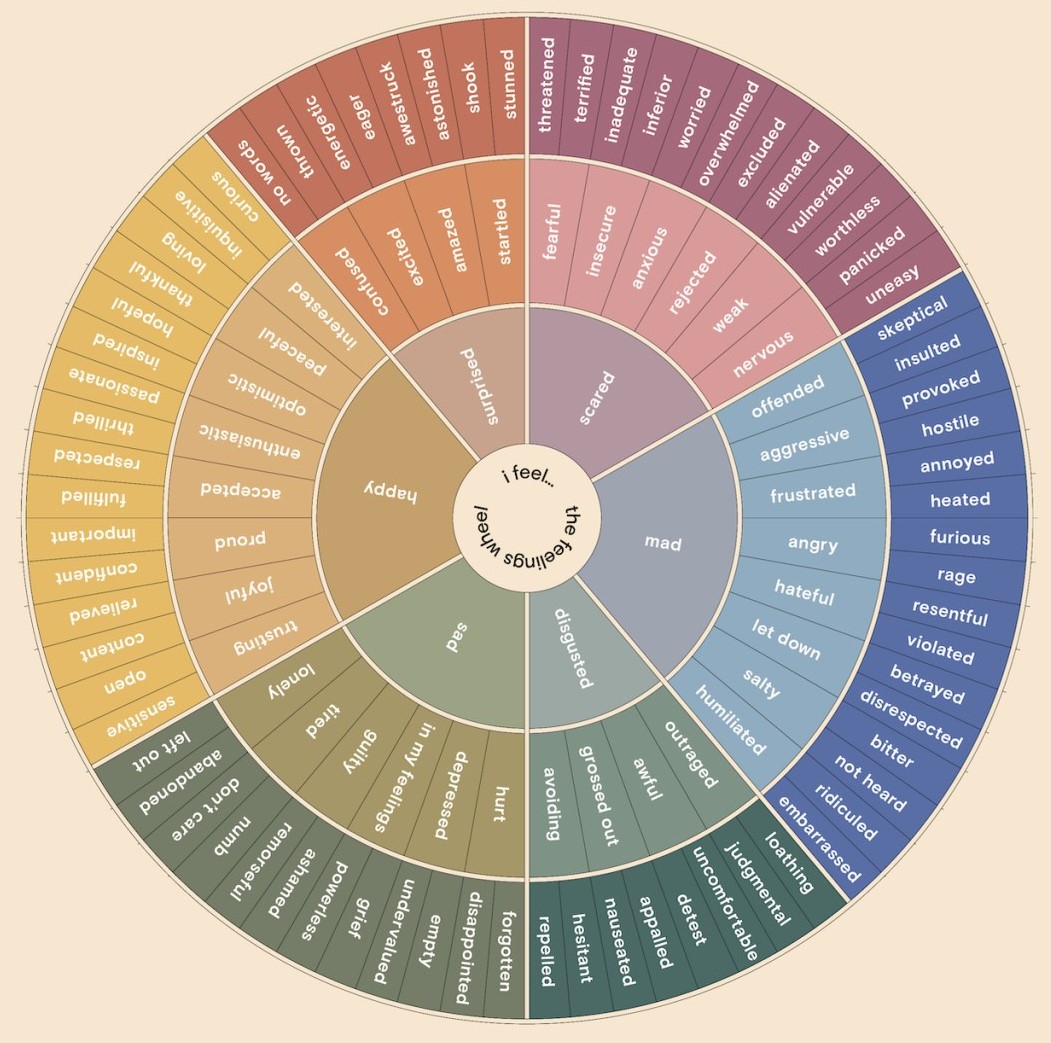}}
        \end{subcaptionbox}
        \begin{subcaptionbox}{W2}[0.26\linewidth]
            {\includegraphics[width=\linewidth]{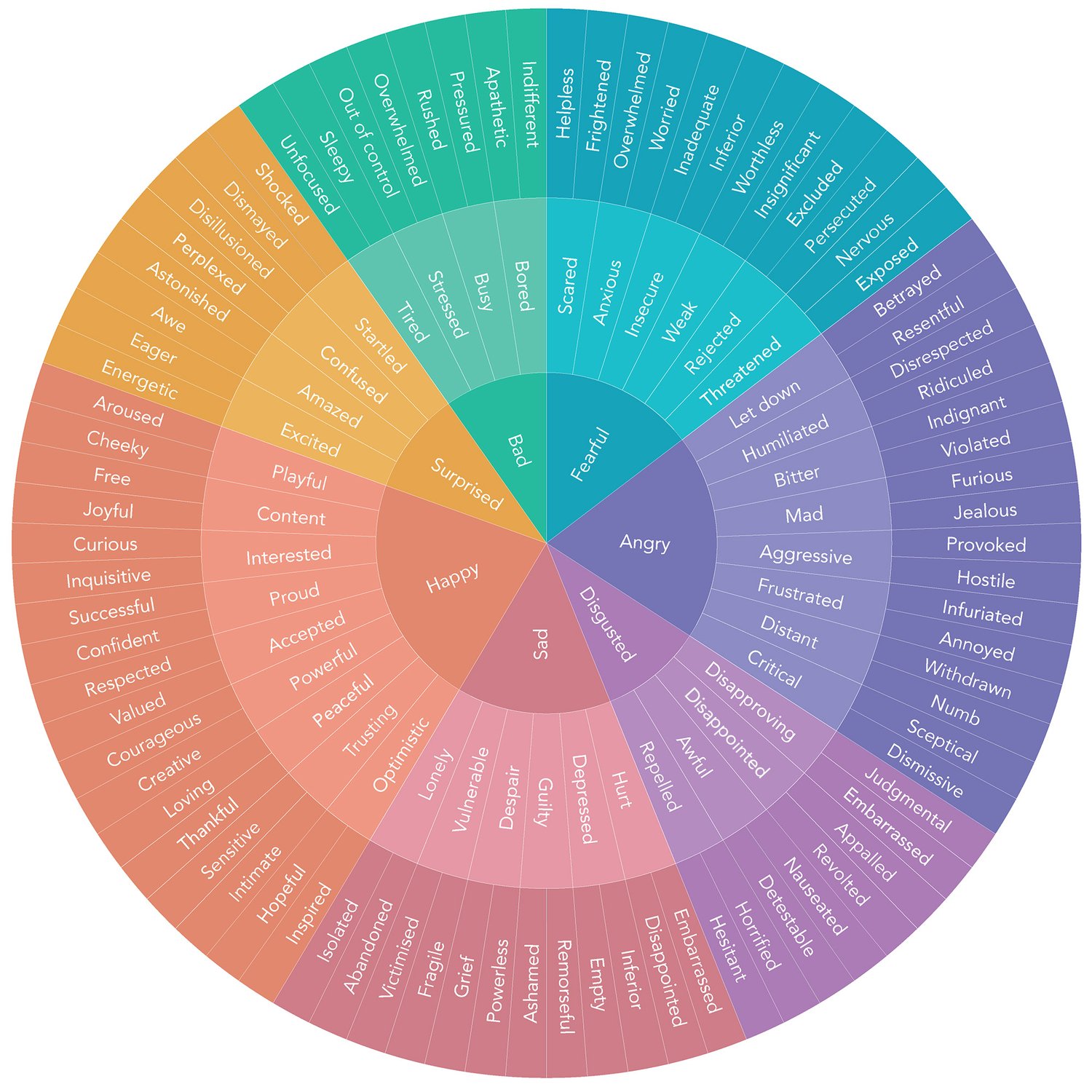}}
        \end{subcaptionbox}
        \begin{subcaptionbox}{W3}[0.26\linewidth]
            {\includegraphics[width=\linewidth]{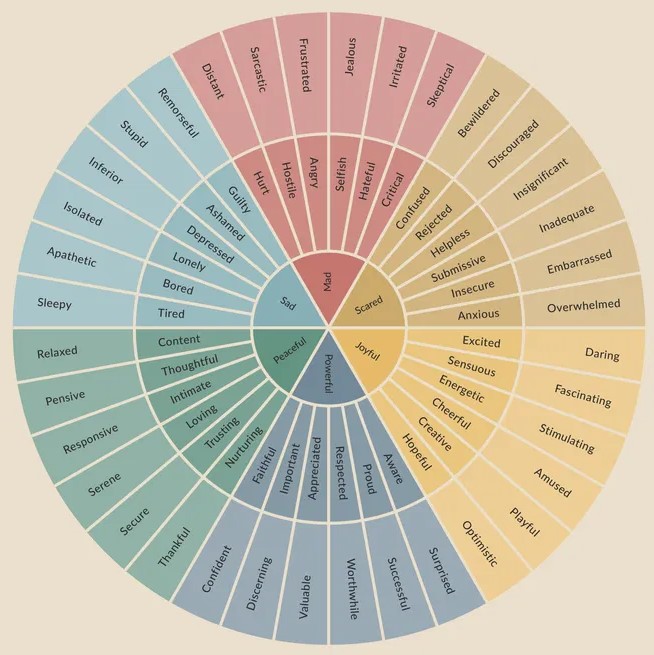}}
        \end{subcaptionbox}
        \begin{subcaptionbox}{W4}[0.26\linewidth]
            {\includegraphics[width=\linewidth]{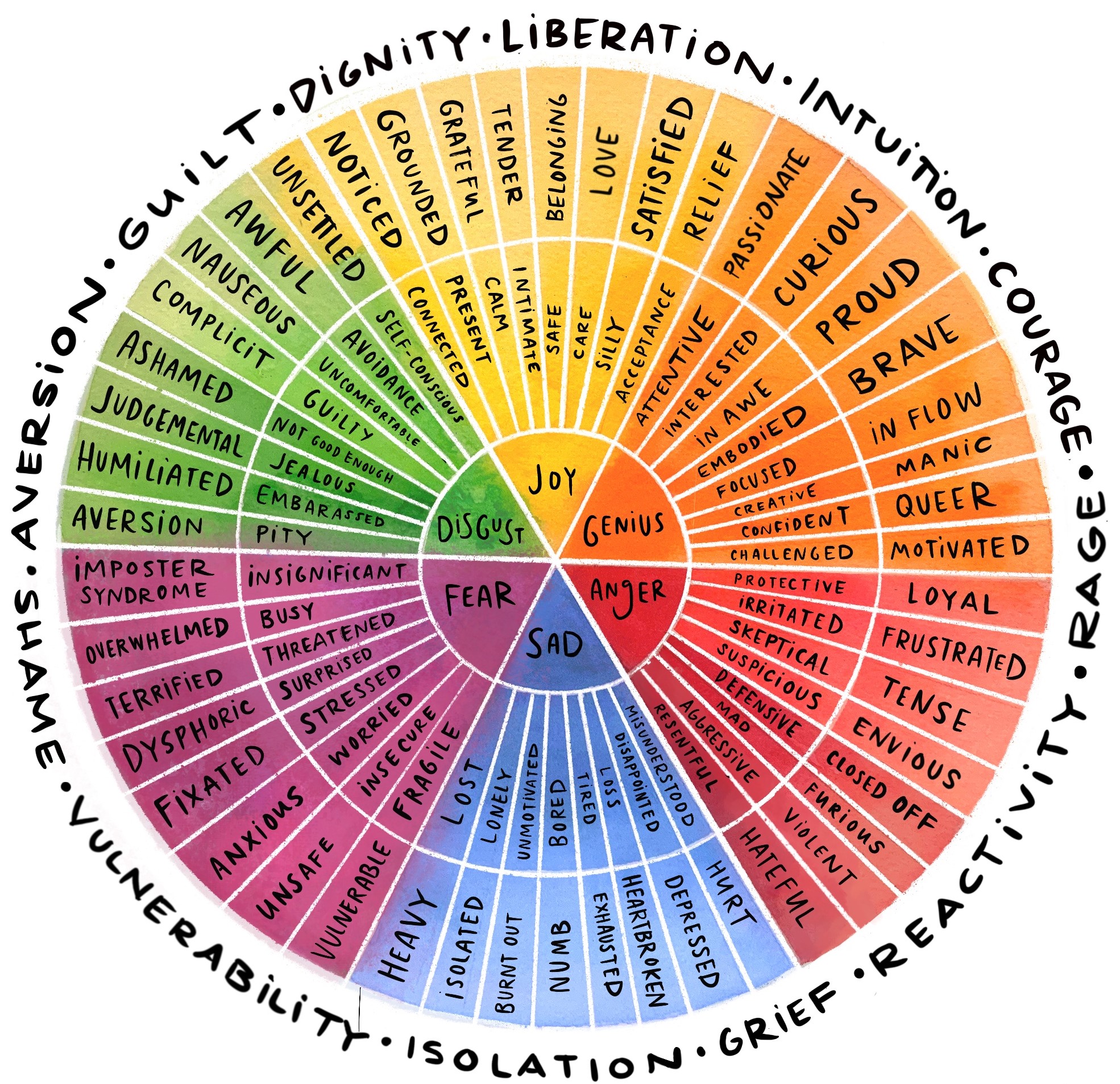}}
        \end{subcaptionbox}
        \begin{subcaptionbox}{W5}[0.26\linewidth]
            {\includegraphics[width=\linewidth]{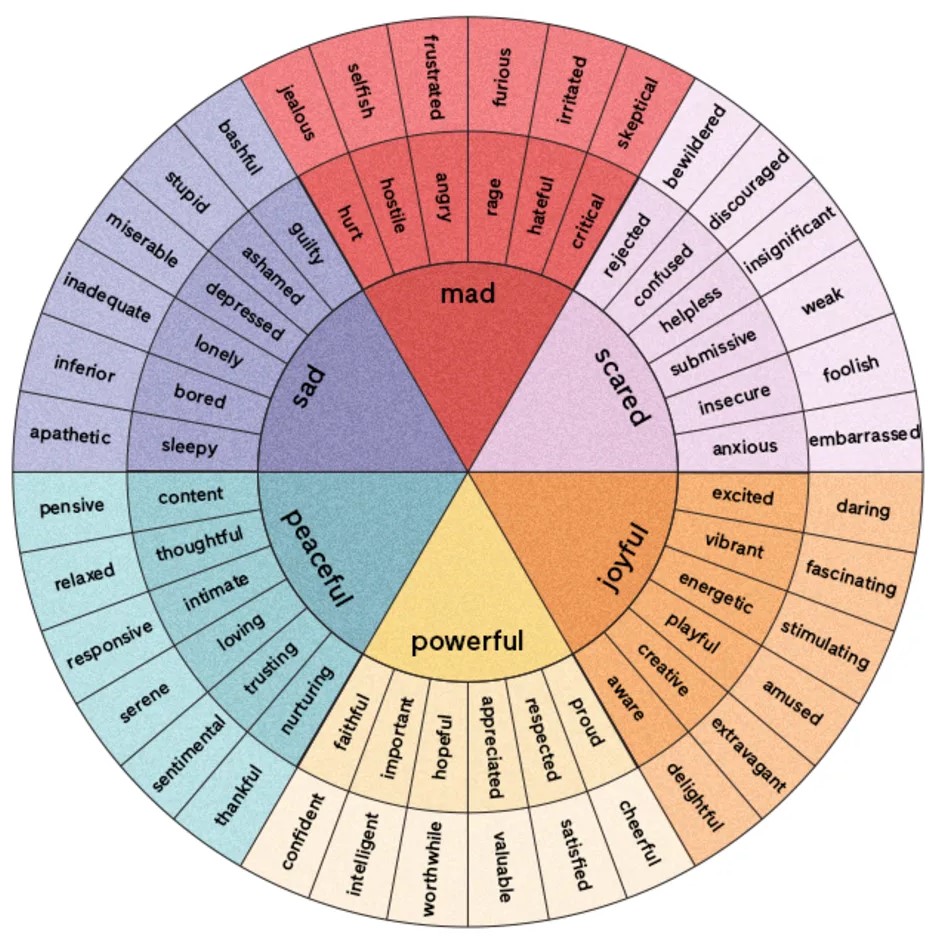}}
        \end{subcaptionbox}
    \end{center}
    \caption{\textbf{Emotion wheels}. This paper uses five emotion wheels to calculate EW-based metrics, ensuring a fair comparison with previous work \cite{lian2025affectgpt,lian2025ov}.}
    \label{fig:ew}
\end{figure}

The emotion wheel is a structured visual tool designed to categorize and identify human emotions. It organizes emotions into core categories (the inner sections of the wheel) and their nuanced variations (the outer sections), providing intuitive insights into emotional structure. This paper uses five emotion wheels to compute EW-based metrics, aligning with prior work \cite{lian2025affectgpt} to ensure fair comparisons. Figure \ref{fig:ew} illustrates these emotion wheels.

\section{EW-based Metric}
\label{appendix:ew_metric}
In this section, we present a detailed calculation formula for EW-based metrics \cite{lian2025ov}. The computation method primarily consists of two components: eliminating the influence of synonyms and defining set-level metrics.

\subsection{Handling Synonyms}
To mitigate the impact of synonyms, we employ a three-level hierarchical grouping strategy:

\textbf{(L1)} We map different forms of words to their base form. For example, \emph{happier} and \emph{happiness} are mapped to \emph{happy}. To achieve this, for a base word $w$, we first use GPT to generate various forms of $w$, with the output denoted as $l_1(w)$. Since different results may be produced across multiple calls, we repeat this process ten times and merge all the generated words, resulting in $\hat{l}_1(w)$. To map these different word forms back to their base form, we simply map $\hat{l}_1(w)$ to $w$, and we refer to this mapping function as $F_{l_1}(\cdot)$.
	
\textbf{(L2)} We map synonyms to a unified word. For example, \emph{happy} and \emph{joyful} are mapped to \emph{happy}. To achieve this, similar to \emph{\textbf{L1}}, we use GPT to generate synonyms for a given word $w$, repeat the process multiple times, and then merge all the generated words, obtaining $\hat{l}_2(w)$. To map the synonyms to their unified form, we map each element in $\hat{l}_2(w)$ to $w$, and we refer to this mapping function as $F_{l_2}(\cdot)$.
	
\textbf{(L3)} The emotion wheel offers natural grouping information for emotions \cite{plutchik1980general}. Since there is no consensus on the emotion wheel, we adopt $K=5$ distinct emotion wheels, consistent with prior work \cite{lian2025affectgpt,lian2025ov} to ensure fair comparisons. Figure~\ref{fig:ew} visualizes all five emotion wheels. For each wheel $w_k$, we map all outer labels to their corresponding inner labels. This function is denoted as $F_{l_3}^{w_k}(\cdot)$.

Finally, the clustering functions can be summarized as:
\begin{equation}
G_{w_k}(\cdot) = F_{l_3}^{w_k}{\left(F_{l_2}\left(F_{l_1}\left(\cdot\right)\right)\right)}, k \in [1, K].
\end{equation}

\subsection{Set-level Metric} 
Since we do not restrict the number of predicted and annotated emotions, we employ set-level evaluation metrics. Specifically, suppose the dataset contains $N$ samples. For a sample $x_i$, the true labels are $\mathbf{Y}_i=\{y_i^j\}_{j=1}^{n_i}$ and the predicted labels are $\mathbf{\hat{Y}}_i=\{\hat{y}_i^j\}_{j=1}^{\hat{n}_i}$. Here, $n_i$ and $\hat{n}_i$ are the number of predicted and annotated emotions, respectively. The evaluation metrics are defined as follows:
\begin{equation}
\mbox{Precision}_{\mbox{s}}^{k} = \frac{1}{N}\sum_{i=1}^{N}\frac{\left|G_{w_k}( \mathbf{Y}_i ) \cap G_{w_k}(\mathbf{\hat{Y}}_i)\right|}{\left|G_{w_k}(\mathbf{\hat{Y}}_i)\right|},
\end{equation}
\begin{equation}
\mbox{Recall}_{\mbox{s}}^{k} = \frac{1}{N}\sum_{i=1}^{N}\frac{\left|G_{w_k}( \mathbf{Y}_i ) \cap G_{w_k}(\mathbf{\hat{Y}}_i)\right|}{\left|G_{w_k}(\mathbf{{Y}}_i)\right|},
\end{equation}
\begin{equation}
\mbox{F}_{\mbox{s}}^{k} = 2\times\frac{\mbox{Precision}_{\mbox{s}}^{k}\times\mbox{Recall}_{\mbox{s}}^{k}}{\mbox{Precision}_{\mbox{s}}^{k}+\mbox{Recall}_{\mbox{s}}^{k}}.
\end{equation}

\textbf{In this set-wise operation, we automatically remove duplicate emotion words.}
Finally, we compute the average F1-score across all emotion wheels as the final score:
\begin{equation}
\text{EW}\left(\mathbf{Y}_i, \mathbf{\hat{Y}}_i\right) = \frac{1}{K}\sum_{k=1}^{K}\mbox{F}_{\mbox{s}}^{k}.
\end{equation}

\section{Details of Alignment Reward}
\label{appendix:alignment}

\subsection{Emotion Word Extraction}
To compute the alignment reward, we need to extract emotion words $e^t$ from $o^t$. To do so, we input $o^t$ into Qwen2.5-7B using the following prompt: \textcolor[rgb]{0.93,0.0,0.47}{\emph{Please assume the role of an expert in the field of emotions. We provide clues related to the emotional state of a character. Based on the provided clues, please identify the character's emotional states. Separate different emotional categories with commas, and output only the clearly identifiable emotional categories in a list format. If no emotions can be identified, please return an empty list.}}

\subsection{Similarity Calculation}
Alignment reward involves computing the similarity between two emotion sets. In practice, we use the \emph{Jaccard Similarity Coefficient}, which measures the similarity between two sets by comparing the size of their intersection to the size of their union. For clarity, we adopt the notations defined in Appendix \ref{appendix:ew_metric}, and the method for calculating the \emph{Jaccard Similarity Coefficient} is provided below:
\begin{equation}
\mbox{Similarity}_{\mbox{s}}^{k} = \frac{1}{N}\sum_{i=1}^{N}\frac{\left|G_{w_k}( \mathbf{Y}_i ) \cap G_{w_k}(\mathbf{\hat{Y}}_i)\right|}{\left|G_{w_k}( \mathbf{Y}_i ) \cup G_{w_k}(\mathbf{\hat{Y}}_i)\right|},
\end{equation}
\begin{equation}
    \mathrm{is\_similar}\left(\mathbf{Y}_i, \mathbf{\hat{Y}}_i\right)=\frac{1}{K}\sum_{k=1}^{K}\mbox{Similarity}_{\mbox{s}}^{k}.
\end{equation}

\section{Baseline Details}
\label{appendix:baseline}
\paragraph{SECap \cite{xu2024secap}.}
This is a speech emotion captioning framework designed to describe speech emotions using natural language. Specifically, it employs HuBERT as the audio encoder to extract speech representations and then utilizes the Bridge-Net to obtain emotion-relevant speech features. Finally, these features are fed into the LLM, which generates natural language captions describing speech emotions.

\paragraph{SALMONN \cite{tang2023salmonn}.}
This model is capable of processing both speech and audio. Specifically, it employs Whisper \cite{radford2023robust} to encode speech and BEATs \cite{chen2023beats} to encode audio. The resulting features are then integrated into the LLM via a window-level Q-Former. Additionally, a LoRA adapter is applied to enhance the LLM’s instruction-following and question-answering capabilities.

\paragraph{Qwen-Audio \cite{chu2023qwen}.}
This model is capable of handling 30 different tasks and a variety of audio types, demonstrating universal audio understanding capabilities. Specifically, it employs an audio encoder to map input audio into hidden features. To address interference between different tasks, it adopts a multi-task input format. The output features are then passed to LLMs for answer generation.

\paragraph{Otter \cite{li2025otter}.}
This model is built upon the OpenFlamingo framework \cite{awadalla2023openflamingo} and trained on the MIMIC-IT dataset \cite{li2025otter}, demonstrating strong capabilities in multimodal perception, reasoning, and in-context learning.

\paragraph{Chat-UniVi \cite{jin2024chat}.}
This is a unified vision-language model capable of processing both images and videos. It employs dynamic visual tokens that capture spatial details in images and temporal relationships in videos.

\paragraph{Video-LLaVA \cite{lin2024video}.}
It aligns images and videos before projection, enabling the LLM to learn from a unified visual representation and equipping it with the ability to comprehend both images and videos simultaneously.

\paragraph{Video-ChatGPT \cite{maaz2024video}.}
This model averages frame-level features across temporal and spatial dimensions, extracting both spatial and temporal video features. These features are then projected into the text token space, enabling the LLM to generate responses.

\paragraph{VideoChat \cite{li2025videochat}.}
It uses video foundation models to encode videos as embeddings, then employs LLMs for response generation. After instruction fine-tuning, the model demonstrates promising capabilities across various video applications.

\paragraph{VideoChat2 \cite{li2024mvbench}.}
This model leverages diverse instruction-tuning data and employs a three-stage progressive multimodal training approach, comprising vision-language alignment, vision-language connection, and instruction tuning. As a result, it outperforms leading models across multiple tasks.

\paragraph{LLaMA-VID \cite{li2024llama}.}
This model represents each image with two tokens: a context token and a content token. The context token encodes the global image information based on user input, while the content token captures visual details within the image. By condensing each image into just a few tokens, the model can efficiently understand and process long videos.

\paragraph{mPLUG-Owl \cite{ye2023mplug}.}
This approach equips LLMs with vision-language capabilities through a two-stage training paradigm, enabling strong performance on instruction-following, visual understanding, and reasoning tasks.

\paragraph{PandaGPT \cite{su2023pandagpt}.}
It uses ImageBind \cite{girdhar2023imagebind} to map multiple modalities into a shared embedding space, then employs a projector to transform these features into the LLM's input space, and finally leverages the LLM to generate responses.

\paragraph{OneLLM \cite{han2024onellm}.}
This model aligns eight modalities with language through a unified framework. Its architecture consists of a universal encoder, a universal projection module, and modality-specific tokens. Additionally, it employs a progressive multimodal alignment pipeline, enabling it to comprehend eight input types and generate responses based on user queries.

\paragraph{R1-Omni \cite{zhao2025r1}.}
This framework also leverages reinforcement learning for emotion recognition. Unlike AffectGPT-R1, which focuses on open-vocabulary emotions, this work targets basic emotions. Due to the different objectives, the reward functions used in R1-Omni differ from those employed in AffectGPT-R1. A more detailed comparison between R1-Omni and AffectGPT-R1 is provided in Appendix~\ref{appendix:novelty}.

\paragraph{Emotion-LLaMA \cite{cheng2024emotion}.}
This framework follows the architecture of traditional MLLMs and consists of three key components: encoders, projectors, and LLMs. Specifically, it inputs text, audio, peak frames, and three levels of visual cues (i.e., local, temporal, and global representations) and employs modality-specific encoders to convert these inputs into corresponding hidden features. These features are then aligned into a shared dimensional space with text embeddings through a linear projection mechanism. Finally, the LLM integrates information from all modalities for emotion recognition and reasoning. Experimental results on benchmark datasets demonstrate that this framework achieves improved performance over previous solutions.

\paragraph{AffectGPT \cite{lian2025affectgpt}.}
This framework differs from previous MLLMs, which typically delegate the entire multimodal fusion process to the language model. Instead, it relocates cross-modal interaction outside the language model and incorporates a pre-fusion operation to enhance multimodal integration, emphasizing the inherently multimodal nature of human emotions. Specifically, the framework introduces two types of pre-fusion mechanisms: Q-Former and attention mechanisms. Both modules are computationally efficient and introduce only a small number of trainable parameters. Experimental results demonstrate the effectiveness of AffectGPT in generalized emotion understanding, including sentiment analysis, basic emotion recognition, and fine-grained emotion detection.

\section{Dataset Details}
\label{appendix:dataset}

\subsection{Training Datasets}

\paragraph{MER-Caption+ \cite{lian2025affectgpt}.}
MER-Caption+ is a large-scale descriptive emotion dataset comprising approximately 31K samples and 2K emotion labels. To reduce annotation costs, it introduces a novel \emph{model-led human-assisted} strategy that leverages human priors to guide both description generation and sample filtering. Specifically, in the description generation stage, the pre-trained models are carefully selected to maximize performance on fine-grained emotion detection. During sample filtering, a two-level filtering process is employed to remove samples with mismatched audio and video, abnormally long descriptions, and those that perform poorly on sentiment analysis and basic emotion recognition.

\paragraph{MER2025-OV \cite{lian2025mer}.}
MER2025-OV is specifically designed for OV-MER. Unlike the automatically annotated MER-Caption+, MER2025-OV is a \emph{high-quality} dataset created through a purely manual annotation process. It employs a multi-round annotation strategy. Initially, a pool of candidate emotions is provided, and four experts in affective computing are asked to select the correct labels and add any missing ones. All selected labels are combined to form the candidate set for the next round. In the second round, four annotators perform another round of labeling, and only labels selected by at least two annotators are retained. This two-stage approach ensures both completeness, by accepting labels chosen by any annotator in the first round, and accuracy, by filtering for consensus in the second. Through this rigorous process, MER2025-OV guarantees the high quality of its annotated emotion labels.

\subsection{Testing Datasets}

\paragraph{OV-MERD+ \cite{lian2025affectgpt}.}
OV-MERD+ is a dataset specifically designed for OV-MER, which is an extended version of OV-MERD \cite{lian2025ov}. The original OV-MERD dataset consists of 332 samples, all of which are fully annotated by humans for open-vocabulary emotions. Following a similar annotation process, OV-MERD+ expands the dataset to a total of 532 samples. In this study, we employ OV-MERD+ as the test set to evaluate the performance of different models on open-vocabulary emotion detection.

\paragraph{MER-UniBench \cite{lian2025affectgpt}.}
\label{appendix:merunibench}
MER-UniBench is a benchmark that encompasses three representative tasks: sentiment analysis, basic emotion recognition, and fine-grained emotion detection. Specifically, sentiment analysis focuses on binary classification into positive and negative sentiments. Basic emotion recognition restricts the label space to a predefined set of basic emotion categories. In contrast, fine-grained emotion detection removes these constraints, allowing predictions across any category. This benchmark is designed to assess generalized emotion understanding capabilities. Table~\ref{tab:merunibench} summarizes the data composition for each task.

\begin{table}[h]
	\centering
	\caption{\textbf{MER-UniBench.} This table presents the data components for three tasks: sentiment analysis, basic emotion recognition, and fine-grained emotion detection. In total, this benchmark includes 12,799 test samples.}
    \label{tab:merunibench}
    \scalebox{0.8}{
        \begin{tabular}{l|c|r|l}
            \toprule
            \textbf{Raw Dataset} & \textbf{Selected Subset} & \textbf{\# Samples} & \textbf{Label Space} \\
            \midrule
            \rowcolor{gray!20}
            \multicolumn{4}{c}{\emph{Sentiment Analysis}} \\
            \midrule
            CMU-MOSI \cite{zadeh2017tensor}      & Test & 686   & \textcolor[rgb]{0.93,0.0,0.47}{\emph{positive, negative}} \\
            CMU-MOSEI \cite{zadeh2018multimodal} & Test & 4,659 & \textcolor[rgb]{0.93,0.0,0.47}{\emph{positive, negative}} \\
            CH-SIMS \cite{yu2020ch}              & Test & 457   & \textcolor[rgb]{0.93,0.0,0.47}{\emph{positive, negative}} \\
            CH-SIMS v2 \cite{liu2022make}        & Test & 1,034 & \textcolor[rgb]{0.93,0.0,0.47}{\emph{positive, negative}} \\
            \midrule
            \rowcolor{gray!20}
            \multicolumn{4}{c}{\emph{Basic Emotion Recognition}} \\
            \midrule
            MER2023 \cite{lian2023mer}           & MER-MULTI & 411   & \textcolor[rgb]{0.93,0.0,0.47}{\emph{worry, happy, neutral, angry, surprised, sad}} \\
            MER2024 \cite{lian2024mer}           & MER-SEMI  & 1,169 & \textcolor[rgb]{0.93,0.0,0.47}{\emph{worry, happy, neutral, angry, surprised, sad}} \\
            IEMOCAP \cite{busso2008iemocap}      & Session5  & 1,241 & \textcolor[rgb]{0.93,0.0,0.47}{\emph{anger, happiness, sadness, neutral}} \\
            MELD \cite{poria2019meld}            & Test      & 2,610 & \textcolor[rgb]{0.93,0.0,0.47}{\emph{anger, joy, sadness, neutral, disgust, fear, surprise}} \\
            \midrule
            \rowcolor{gray!20}
            \multicolumn{4}{c}{\emph{Fine-grained Emotion Detection}} \\
            \midrule
            OV-MERD+ \cite{lian2025affectgpt}  & All & 532 & \textcolor[rgb]{0.93,0.0,0.47}{\emph{unrestricted label space}} \\
            \bottomrule
        \end{tabular}
    }
\end{table}

\subsection{Data Leakage Prevention and Overlap Removal}
To prevent data leakage, we perform a strict check to ensure that there is no overlap between the training and testing data. During this process, we identify 200 overlapping samples between MER2025-OV (part of the training set) and OV-MERD+ (part of the test set). Consequently, we remove these overlapping samples from MER2025-OV, resulting in 1,000 samples available for training (see Table \ref{tab:dataset}). Apart from this case, we do not observe any other instances of overlap between the training and testing data.

\subsection{Task Clarification}
\label{appendix:task}
In this paper, we focus on \emph{utterance-level} emotion recognition, as opposed to \emph{dialogue-level} emotion recognition. The key difference between these tasks lies in whether conversational history is taken into account. For \emph{utterance-level} emotion recognition, the input consists of a single video clip centered on one character \cite{tsai2019multimodal}. In contrast, \emph{dialogue-level} emotion recognition requires leveraging additional dialogue information to perform emotion inference \cite{poria2017context}. This distinction has led to the development of two types of benchmarks: the \emph{utterance-level} benchmark \cite{lian2026merbench} and the \emph{dialogue-level} benchmark \cite{poria2019emotion}.

In MER-UniBench, although certain datasets (e.g., MELD and IEMOCAP) contain dialogue information, others (e.g., MER2023 and CH-SIMS) do not. Furthermore, even when dialogue context is available in datasets like MELD and IEMOCAP, its inclusion is not strictly necessary and depends on the specific experimental setup \cite{tsai2019multimodal, poria2017context}. Consequently, MER-UniBench is designed as an \emph{utterance-level} benchmark, and this work adopts the same setting to ensure fair comparison with prior studies \cite{lian2025affectgpt}. Dialogue-level emotion recognition is beyond the scope of this paper and will be explored in our future work.

\end{document}